\documentclass[12pt]{paper}
\pdfoutput=1
\usepackage{epsfig}
\usepackage{amsmath}

\newcommand\POWHEG{{\tt POWHEG}}
\newcommand\POWHEGBOX{{\tt POWHEG BOX}}

\newcommand\PYTHIA{{\tt PYTHIA}}
\newcommand\pt{p_{\sss \rm T}}
\newcommand\pT{p_{\sss \rm T}}
\newcommand\kt{k_{\sss \rm T}}
\newcommand\sss{\mathchoice%
{\displaystyle}%
{\scriptstyle}%
{\scriptscriptstyle}%
{\scriptscriptstyle}%
}

\begin{document}

\title{Generation cuts and Born suppression in POWHEG}
\author{
{\bf Paolo Nason}\\
{\small \ \ \ \ INFN, Sezione di Milano-Bicocca,
  Piazza della Scienza 3, 20126 Milan, Italy}\\
{\small \ \  \ \ E-mail: Paolo.Nason@mib.infn.it}\\
\\
{\bf Carlo Oleari}\\
{\small \ \ \ \ Universit\`a di Milano-Bicocca and INFN, Sezione di Milano-Bicocca}\\
{\small \ \ \ \    Piazza della Scienza 3, 20126 Milan, Italy}\\
{\small \ \ \ \    E-mail: Carlo.Oleari@mib.infn.it}
}

\maketitle

\section{Introduction}

In the shower Monte Carlo framework, in processes that already have collinear
singularities at the Born level, a generation cut is needed in order to build
useful event samples. In dijet production, for example, one chooses a minimum
transverse momentum for the $2 \rightarrow 2$ parton process. It is assumed,
and in fact it must be checked, that, in the final showered sample, the
fraction of events that are near the generation cut at the Born level and
that pass the required analysis jet cuts after shower is actually
negligible. In other words, the final results should be insensitive to the
generation cut.  Alternatively, one can introduce a Born suppression
factor. This is a function of the Born kinematics that multiplies the Born
cross section, suppressing its low transverse-momentum region. In this way,
the Born cross section times the suppression factor is an integrable
function, and one can generate events with a probability proportional to this
product. Events are then generated with an associated weight, equal to the
inverse of the Born suppression factor.

The use of a Born suppression factor has the advantage that, for sufficiently
large samples, it yields a correct result. In this case, events that have
particularly low transverse momenta at the Born level, and that after shower
do exhibit hard jets, will appear with a very large weight, equal to the
inverse of the suppression factor. They should be rare enough so as not to
spoil the statistical errors in the results.

It has been observed that, in dijet production~\cite{Alioli:2010xa}, these
events with large weight do in fact cause problems. In physical distributions
large spikes appear from time to time, and it is very difficult to get rid of
them by increasing the statistics.

\section{Origin of spikes in dijet production}

An important mechanism leading to events with large weight in dijet
production was recently identified.

The problem is due to the treatment of the $q \rightarrow q g$ and $g
\rightarrow q \bar{q}$ splittings in the \POWHEGBOX{}. In order to minimize
the number of generated configurations, the \POWHEGBOX{} always generates the
$q \rightarrow q g$ and $g \rightarrow q \bar{q}$ configurations, and never
generates the $q \rightarrow g q$ and $g \rightarrow \bar{q} q$ ones.  The
transverse momentum in final state radiation is defined to be
\begin{equation}
\label{eq:ptdef}
  \pT^2 = 2 E^2  \left( 1 - \cos \theta \right),
  \end{equation}
where $\theta$ is
the angle between the splitting partons, and $E$ is the energy of the emitted
parton (i.e.~parton $k$ in the $i \rightarrow j k$ splitting), both in the
partonic centre of mass. Notice that, if the final-state partons are back to
back, $\pT^2$ is large. Furthermore, $\pT^2$ is large also if the recoiling
parton has small energy, and relatively large angle $\theta$. This region,
however, has small impact in the \POWHEGBOX. If the emitter is a quark, a
soft quark yields no infrared singularities, and thus the region of soft
quarks is power suppressed. If the emitter is a gluon, the \POWHEGBOX{}
suppresses this region with a factor of the form
\begin{equation}
 \label{eq:suppfact}
  \frac{E_{\rm em}^p}{E_{\rm em}^p + E^p},
\end{equation}
where $p$ is a positive number (usually 2) (the emitter is a gluon only if
also the emitted parton is a gluon in the \POWHEGBOX).

In case of jet production, when using the Born suppression factor instead of
a generation cut, the above scheme can yield to events with large weights and
large transverse momenta after showering. They are produced as follows: an
underlying Born event is produced with very small transverse momentum,
corresponding to a $2 \rightarrow 2$ parton scattering at very small angle.
Because of the Born suppression factor, these events are rarely produced, and
have a large weight (proportional to the inverse of the suppression factor).
Suppose now that a splitting process takes place, where a final-state parton,
for example, a gluon, splits into two partons $q \bar{q}$, with $\bar{q}$
carrying most of the energy of the incoming gluon, and $q$ is soft and at a
large angle. This event is phase-space suppressed. However, since the
splitting pair has a small mass, its matrix element has no further
suppression. According to the definition of the radiation transverse momentum
in the final states in the \POWHEGBOX{}, this event has large transverse
momentum (see eq.~(\ref{eq:ptdef})).  When passing the event to \PYTHIA,
further jets with relatively large transverse momentum may be produced. This
event would pass the jet cuts, and have a large weight.

\section{Fixing the problem}
A patch that avoids this problem has been implemented in the SVN
revision~2169.  In order to activate the fix, the user should put the line
\mbox{\tt doublefsr 1} in the {\tt powheg.input} file. If not present, or
different from 1, the program behaves exactly as before.

In the \mbox{\tt doublefsr 1} mode the program does the following:
\begin{itemize}
  \item Considers all splitting processes, including \ $q \rightarrow g q$ and
  $g \rightarrow \bar{q} q$, and not only the  \ $q \rightarrow q g$ and
  $g \rightarrow  q\bar{q} $.
  
  \item Suppresses all splitting processes with a factor proportional to
    eq.~(\ref{eq:suppfact}). Thus both processes $i \rightarrow j k$ and $i
    \rightarrow k j$ are present, with suppression
  \begin{equation}
    \frac{E_j^p}{E_j^p + E_k^p}, \qquad {\rm and} \qquad  \frac{E_k^p}{E_j^p + E_k^p},
  \end{equation}
  respectively. Since the two splitting processes are equivalent, and the sum
  of the two suppression factors is one, one gets back the correct result.
\end{itemize}
The above prescription is formally correct and avoids the problem of the
original \POWHEGBOX{} scheme. Notice that the original prescription was not
incorrect. However, it turned out to be not practical for a large number of
events, where spikes were showing up, and too much statistics would have been
required to smooth them away.

\section{Modification of the {\tt scalup} prescription}\label{sec:chscalup}

We have also studied in the past a modification of the {\tt scalup}
assignment that gets rid of the spike problem. Rather than using the {\tt
  scalup} value provided by the \POWHEGBOX{}, we suggested to recompute its
value as follows. We go to the partonic centre-of-mass frame of the Les
Houches event, and compute the smallest transverse momentum of each final
state parton with respect to the incoming beams and with respect to each
others, and set {\tt scalup} to this value.  The relative squared transverse
momentum of two final-state partons $j$ and $k$ is defined as
\begin{equation}
\pt^2= \left(p_j\cdot p_k\right) \, \frac{E_j E_k}{(E_j+E_k)^2}\;.
\end{equation}

\section{Effects in the output}
We have considered the LHC at 7 TeV.  We have generated two samples of 6M Les
Houches events with the new version of the \POWHEGBOX{} {\tt Dijet} program,
one with the {\tt doublefsr} flag set to 1 (the D1 sample from now on), and
one with no {\tt doublefsr} (the D0 sample).  We have showered these samples
with \PYTHIA{}~6~\cite{Sjostrand:2006za}, keeping the hadronization turned
off. Our \PYTHIA{} settings were:
\begin{verbatim}
c Perugia tune
         CALL PYTUNE(320)
c No QED radiation off quarks
         MSTJ(41)=11
c Hadronization off
         MSTP(111)=0
c Primordial kt off
         MSTP(91)=0
c No multiparton interactions
         MSTP(81)=20
\end{verbatim}
We have used the CTEQ6M~\cite{Pumplin:2002vw} parton densities.

We have also considered the alternative {\tt scalup} definition of
sec.~\ref{sec:chscalup}, that will be referred to as the RS (for
recomputed {\tt scalup}) choice in the following.

In all the figures, two curves are shown, a red and a green one. In the lower
panel, the ratio (displayed in red) of the red over the green is shown.

\begin{figure}[htb]
\begin{center}
\epsfig{file=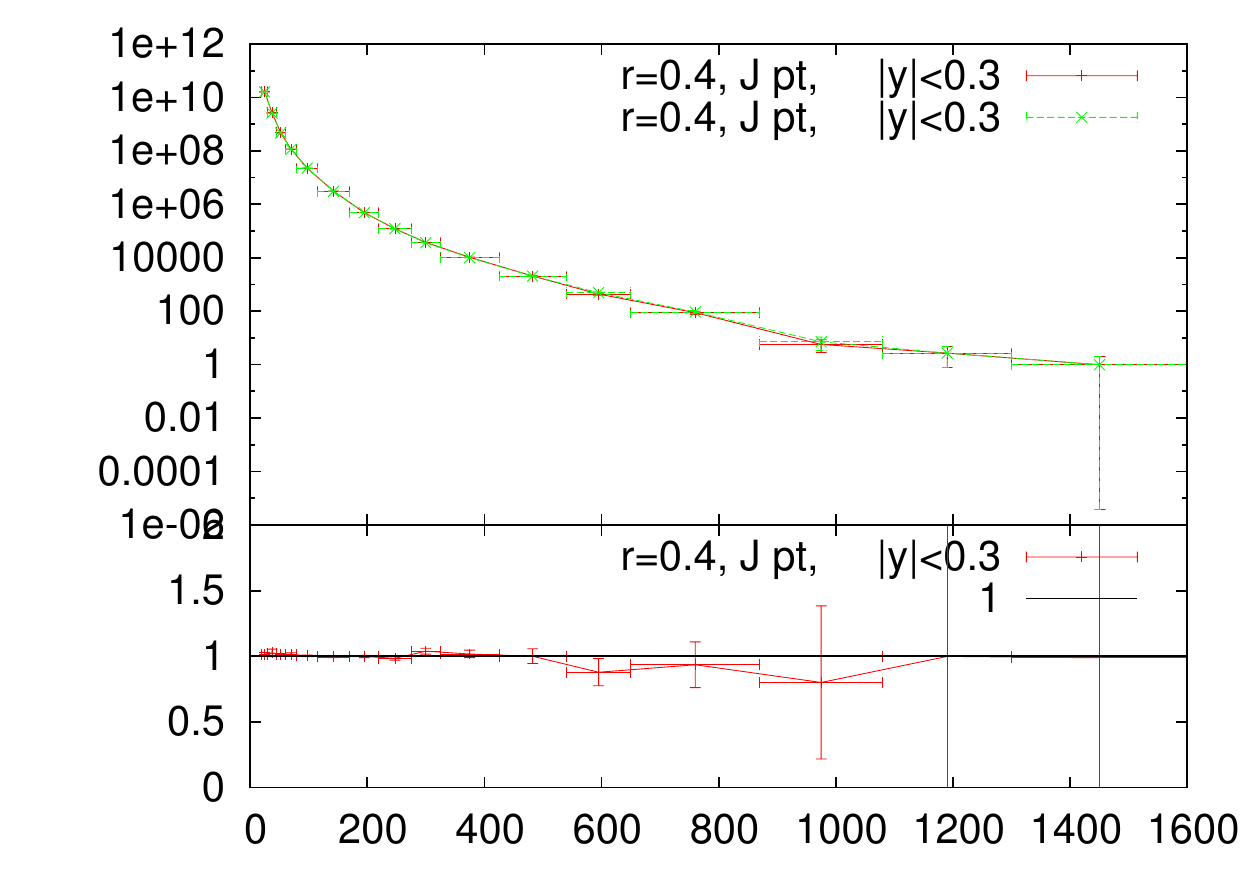,width=0.48\textwidth}
\epsfig{file=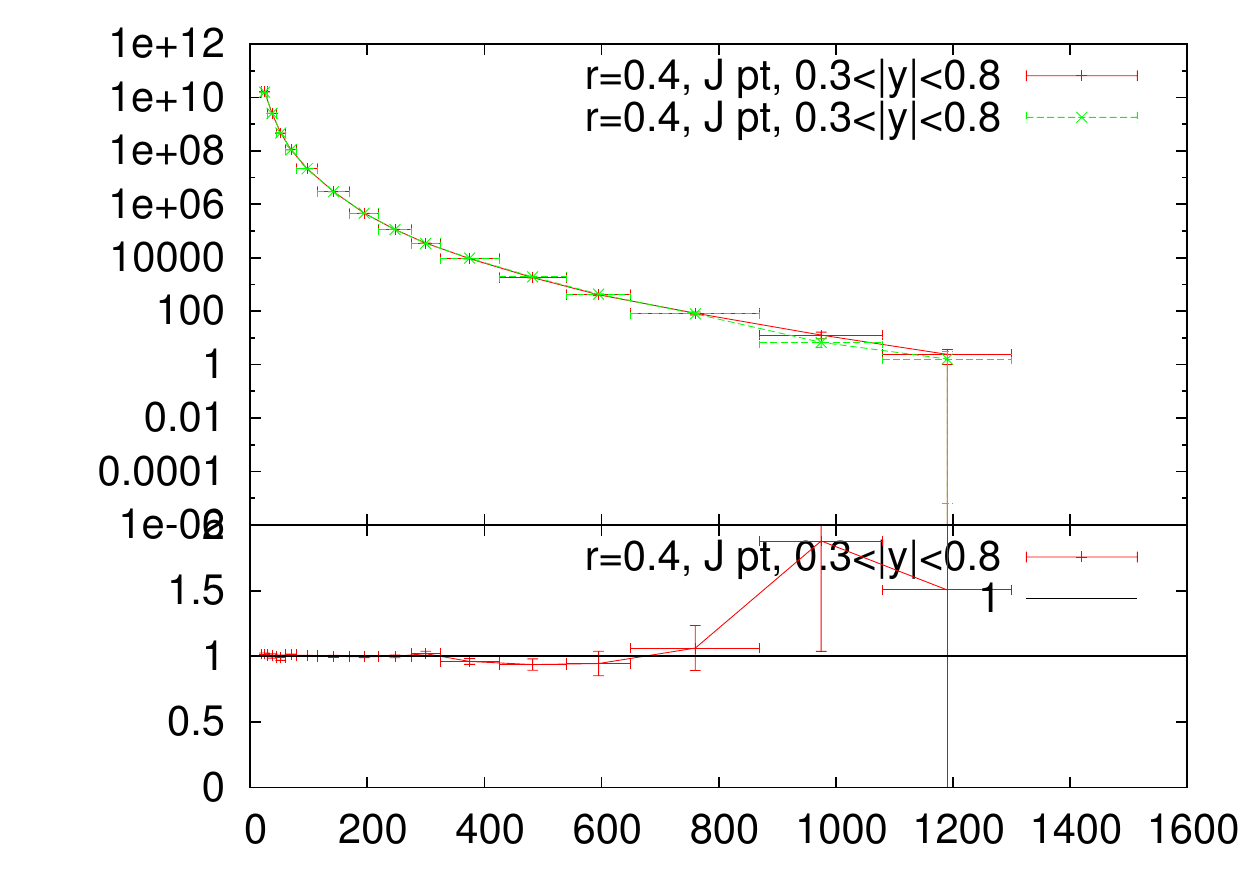,width=0.48\textwidth}
\epsfig{file=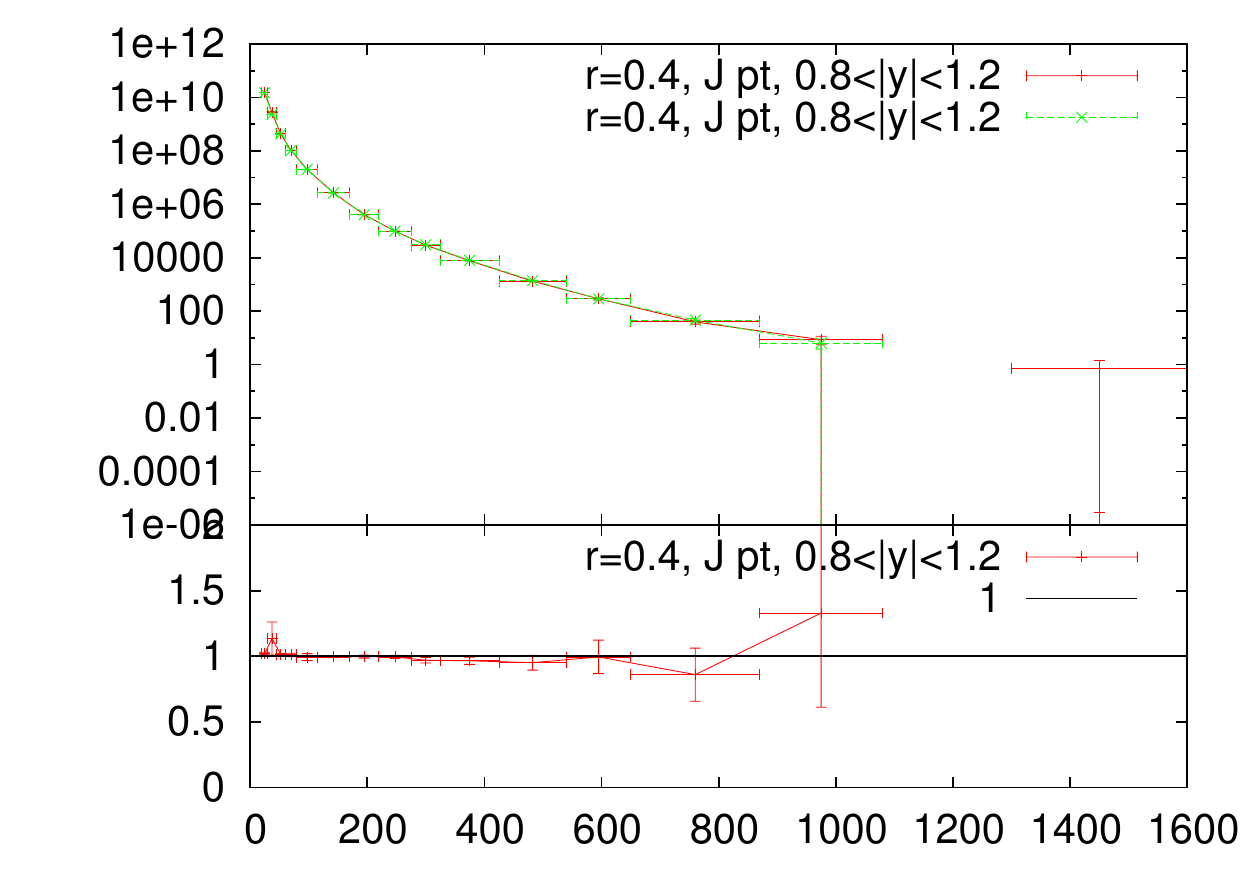,width=0.48\textwidth}
\epsfig{file=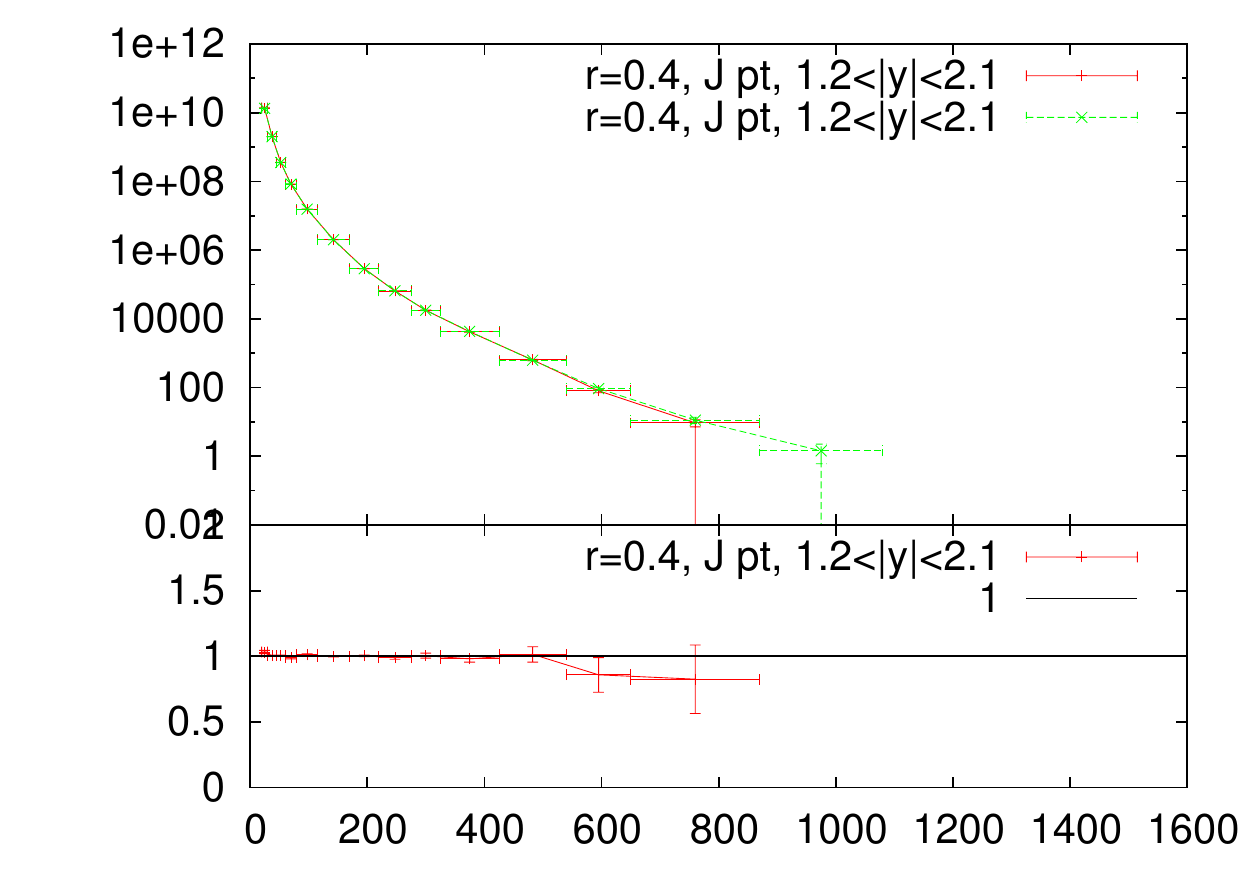,width=0.48\textwidth}
\epsfig{file=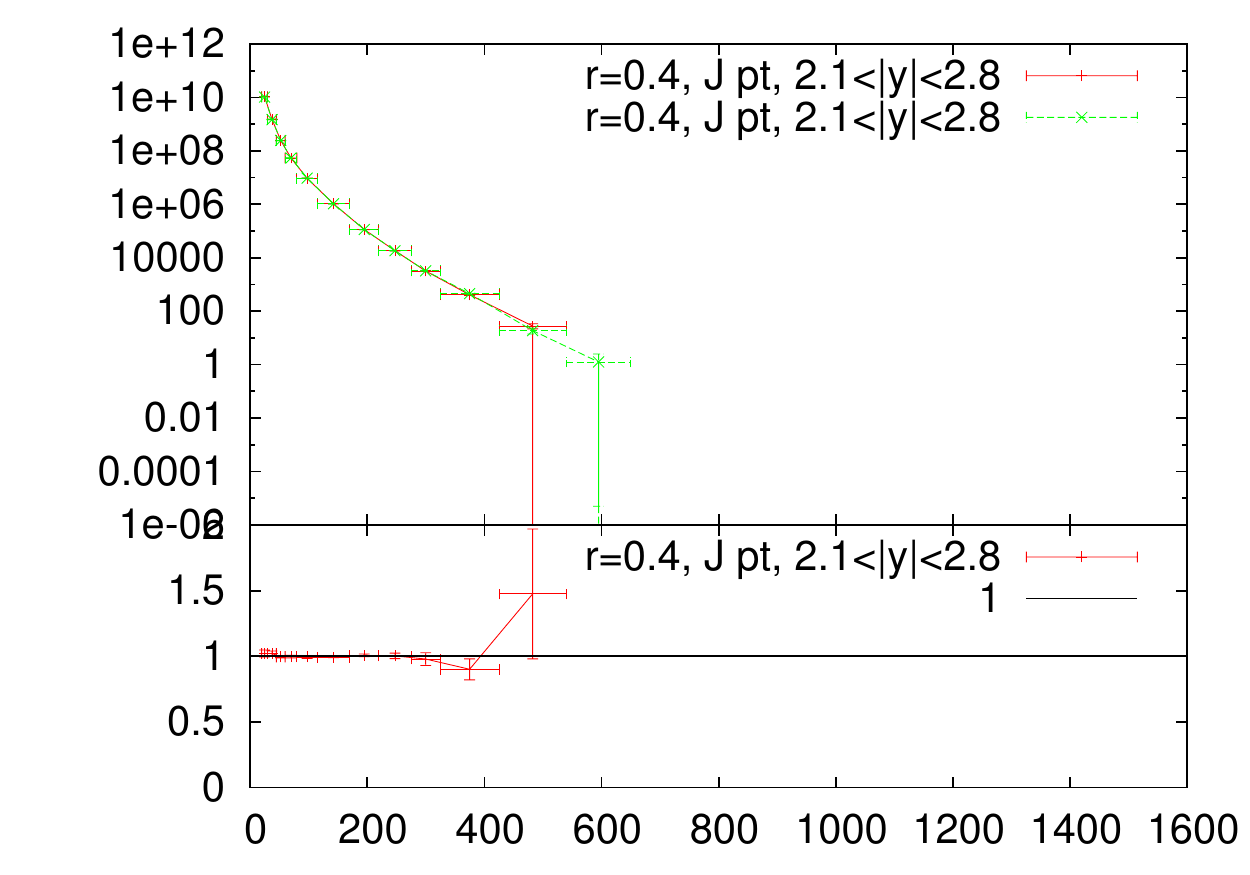,width=0.48\textwidth}
\epsfig{file=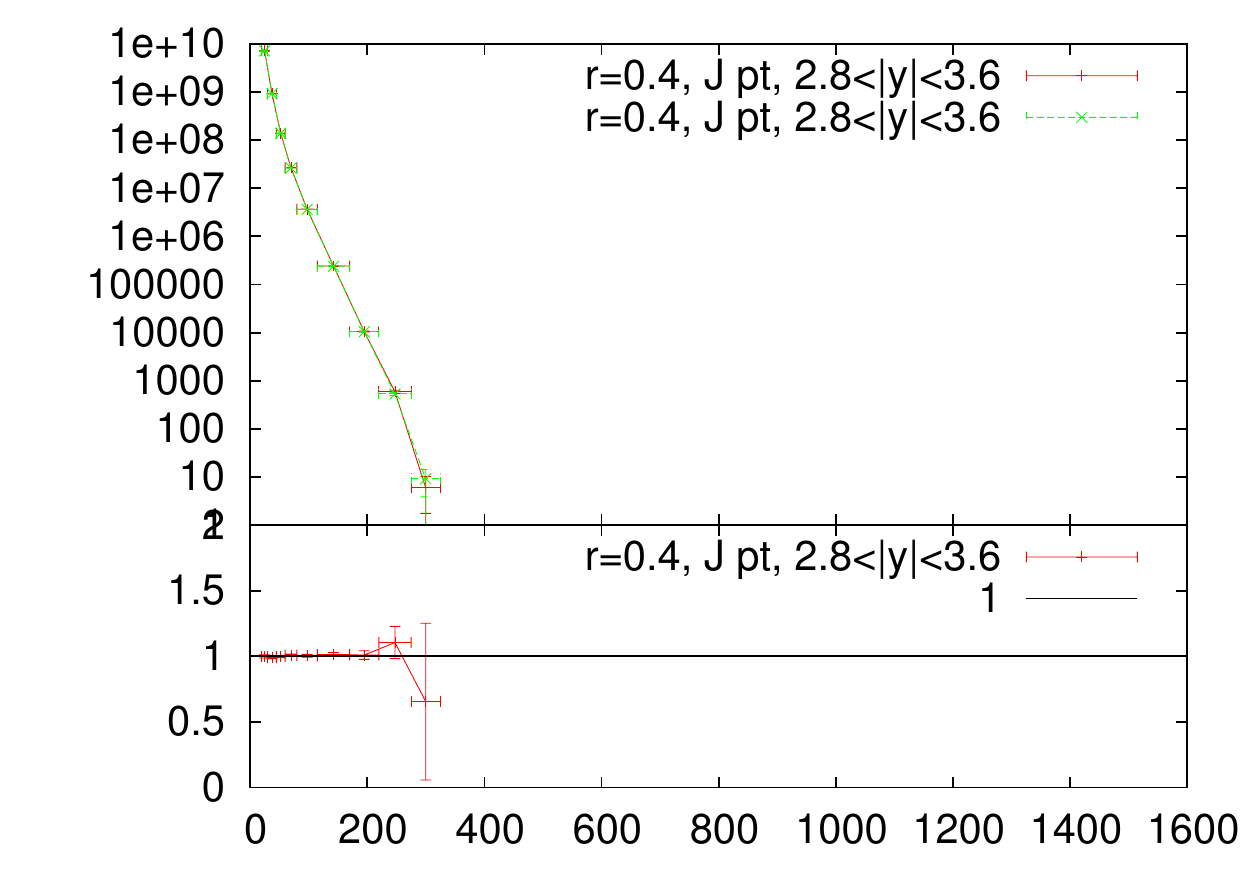,width=0.48\textwidth}
\end{center}
\caption{Comparison of the inclusive jet $\pt$ distribution in the D1 (red)
  and D0 (green) samples, at the Les Houches level (i.e.~before shower).}
\label{fig:pt1}
\end{figure}
We begin by comparing samples D1 and D0 at the level of Les Houches events
(i.e.~before shower). In fig.~\ref{fig:pt1} we show the inclusive
transverse-momentum distributions at various rapidity intervals, for jet
production. Jets are built with the anti-$\kt$
algorithm~\cite{Cacciari:2008gp}, with $R=0.4$.  One can notice the very good
agreement between the two samples.

\begin{figure}[htb]
\begin{center}
\epsfig{file=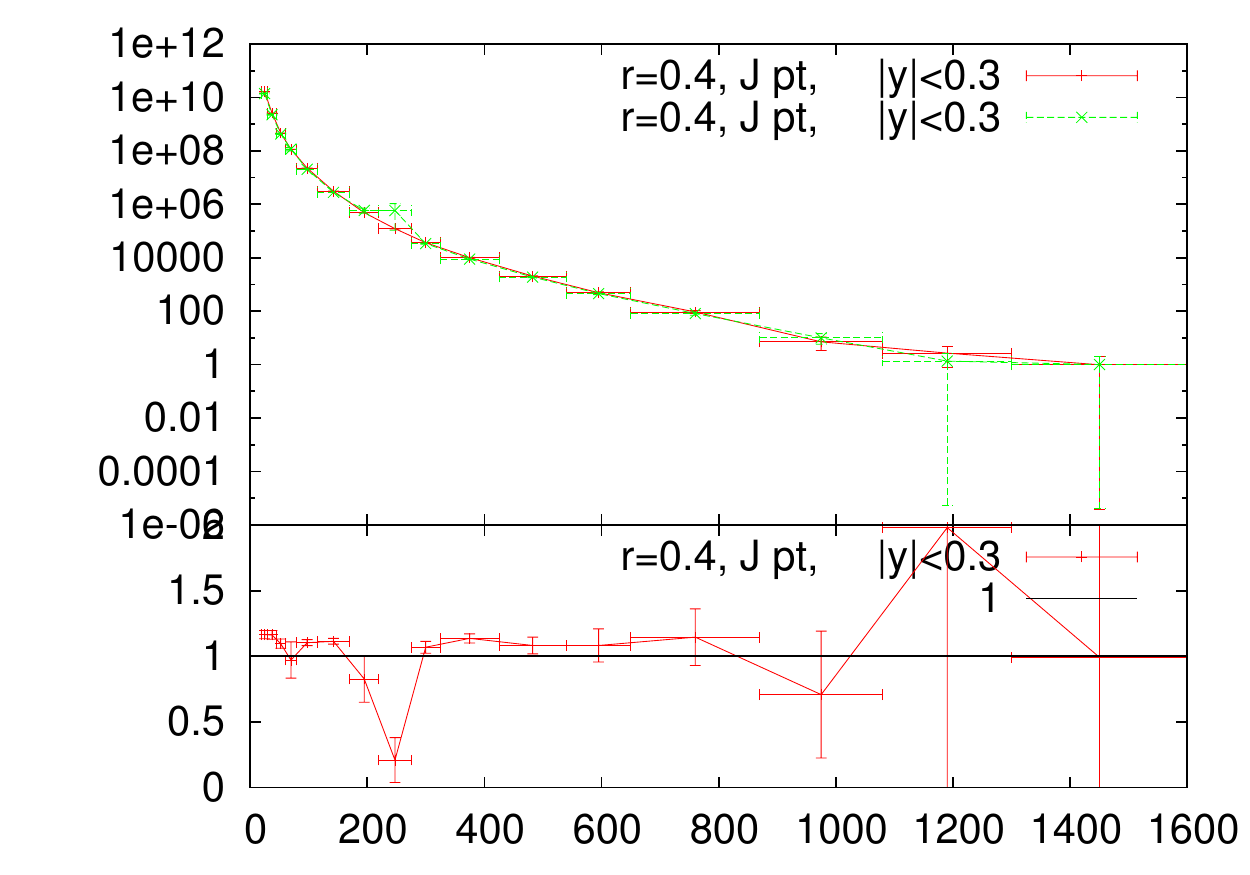,width=0.48\textwidth}
\epsfig{file=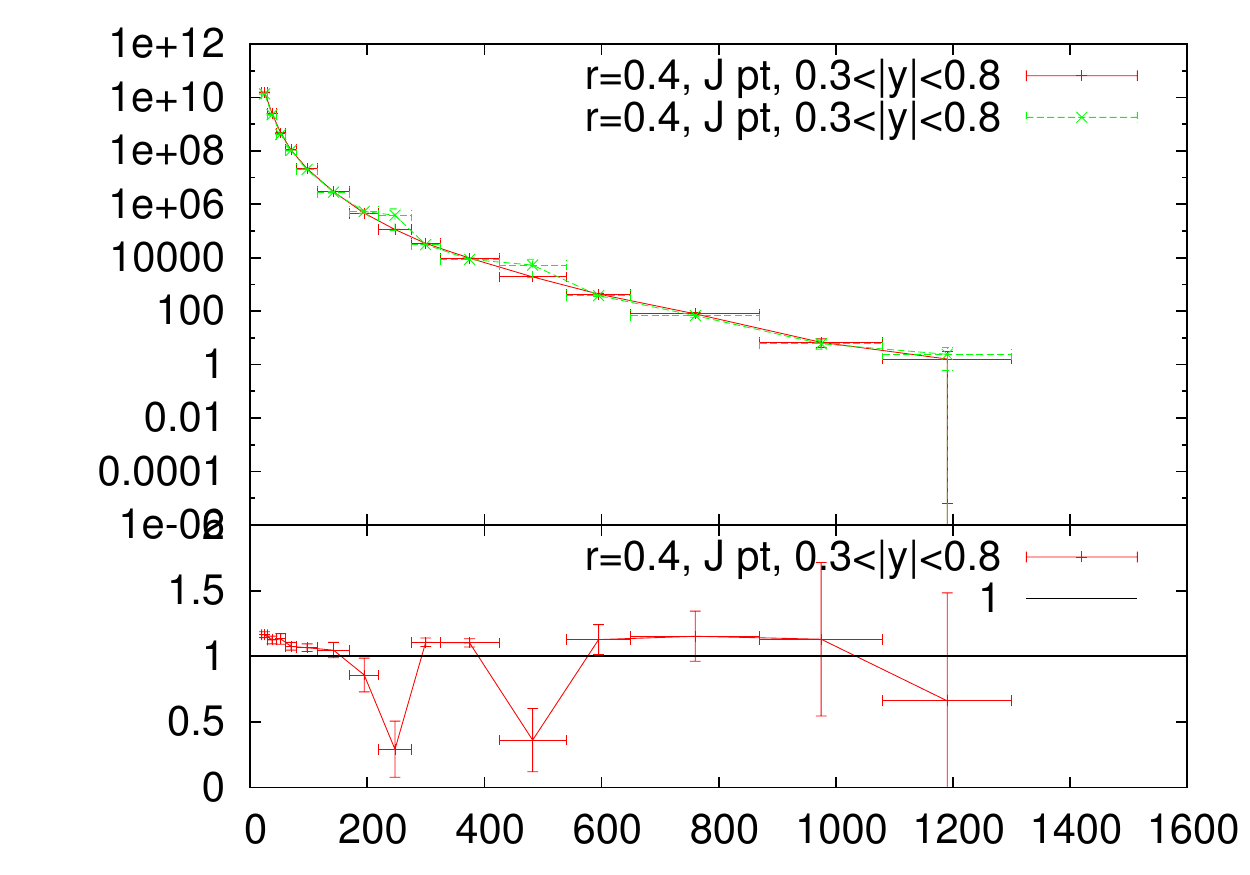,width=0.48\textwidth}
\epsfig{file=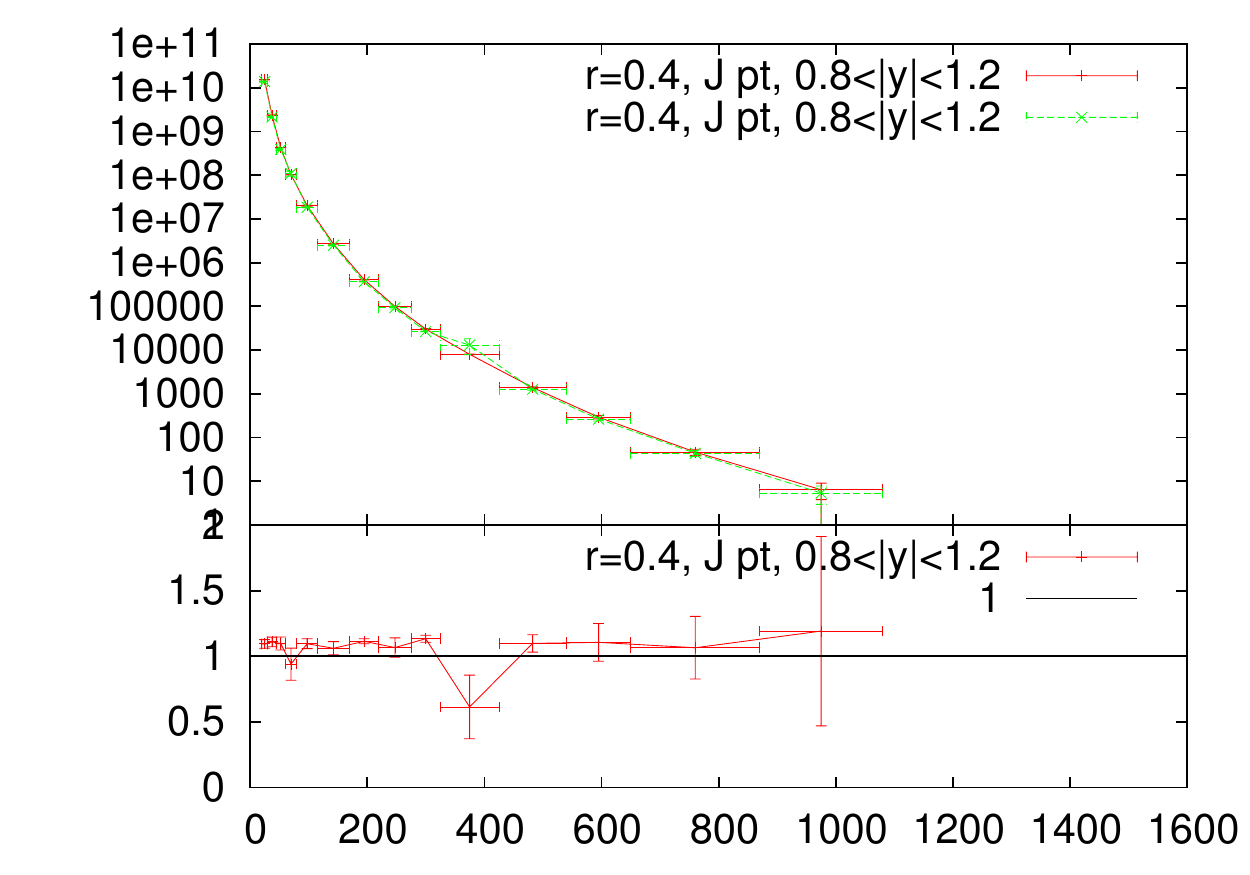,width=0.48\textwidth}
\epsfig{file=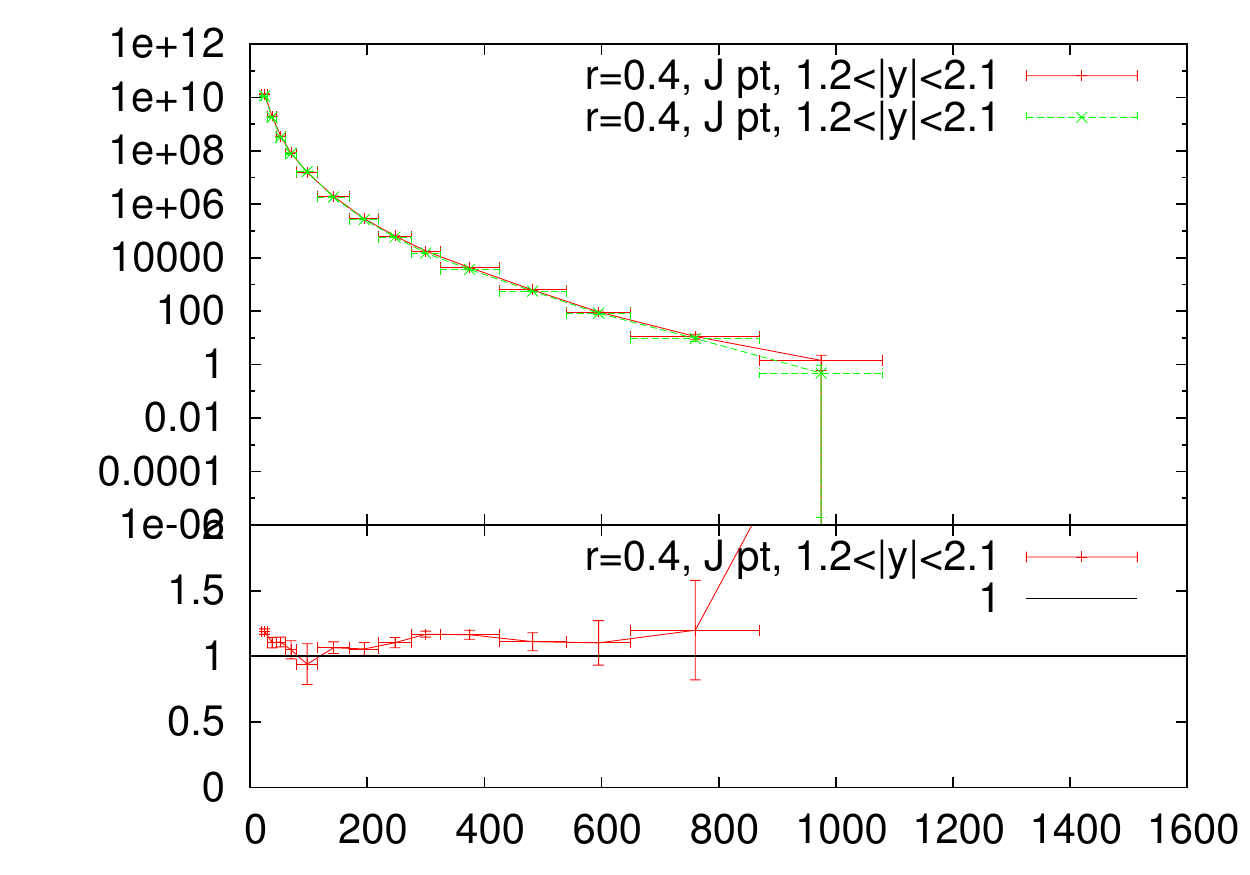,width=0.48\textwidth}
\epsfig{file=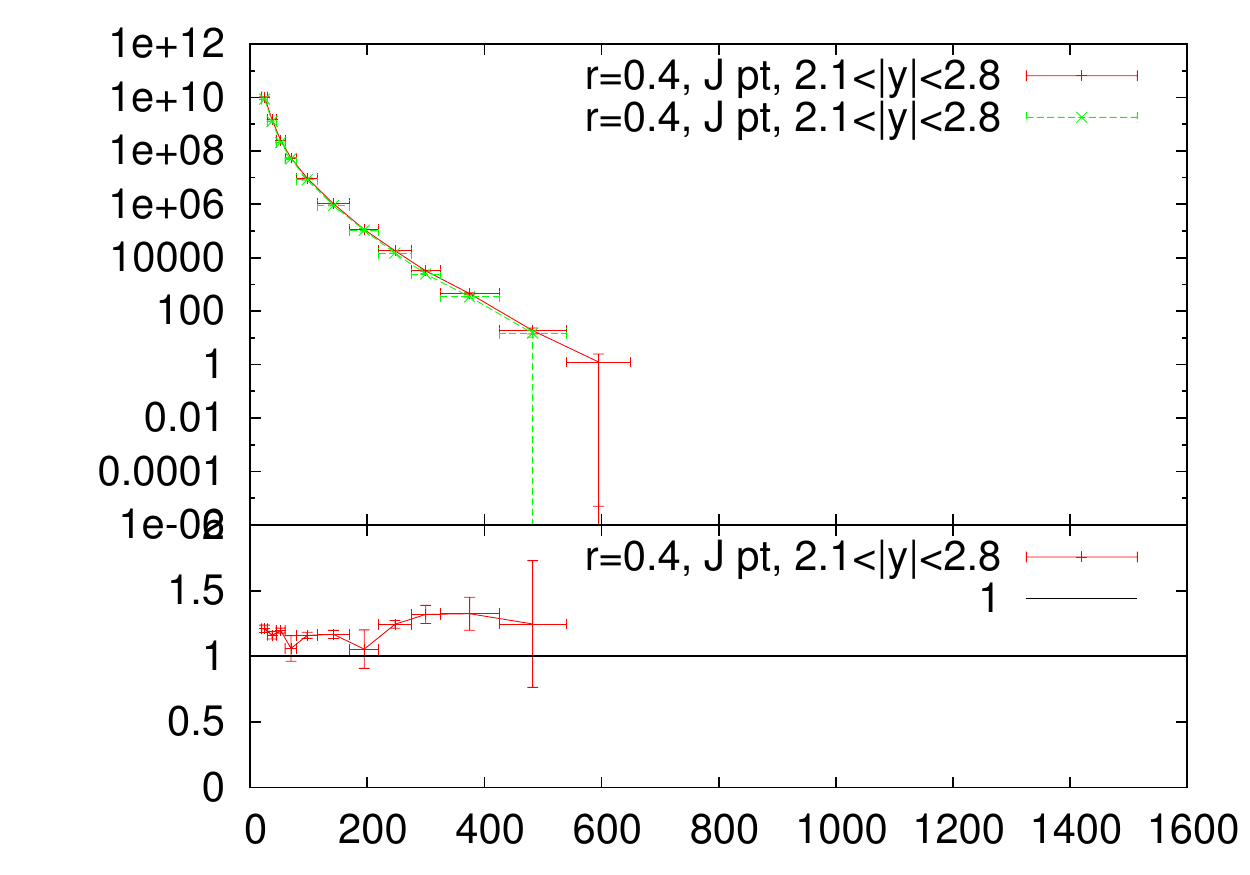,width=0.48\textwidth}
\epsfig{file=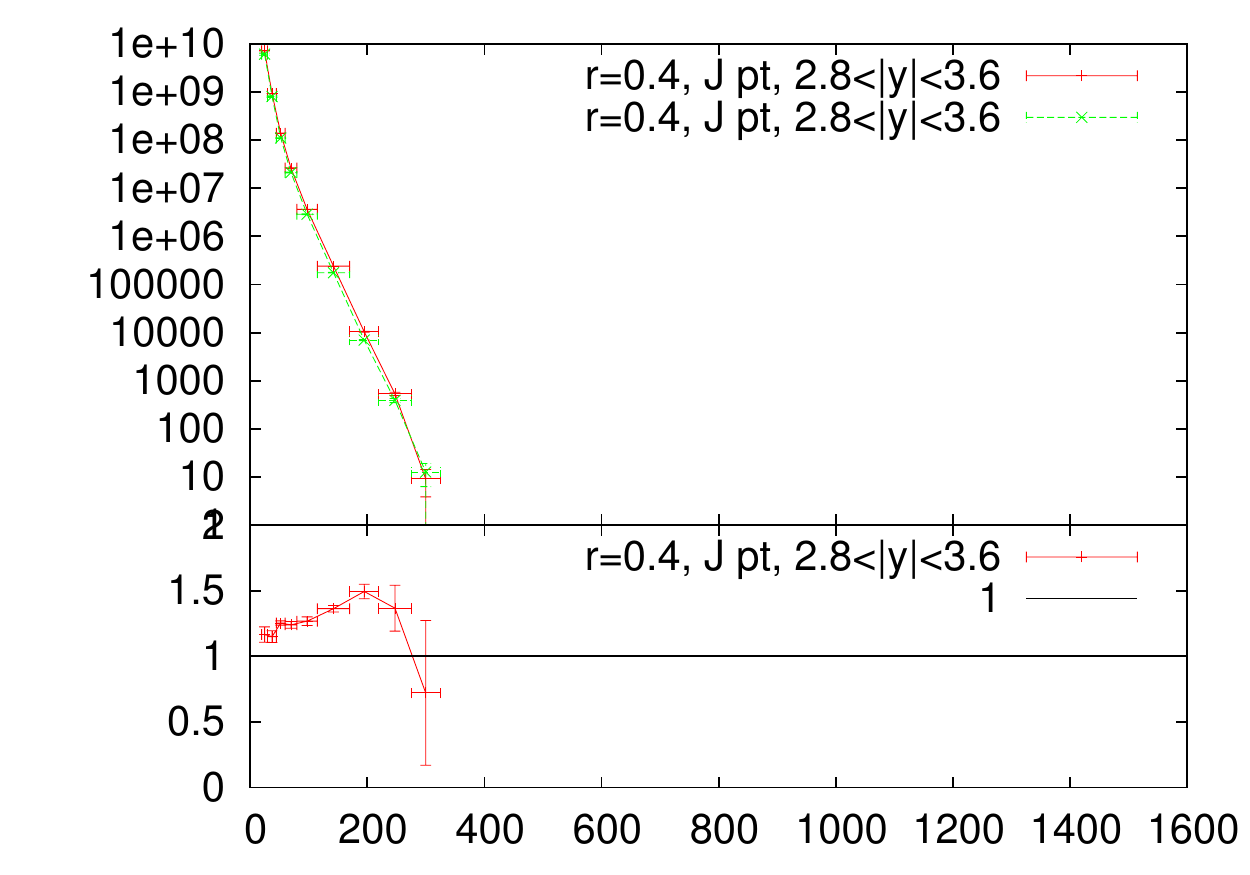,width=0.48\textwidth}
\end{center}
\caption{Comparison of the inclusive jet $\pt$ distribution in sample D0, at
  the Les Houches level (red), and after shower (green).}
\label{fig:pt2}
\end{figure}
\begin{figure}[htb]
\begin{center}
\epsfig{file=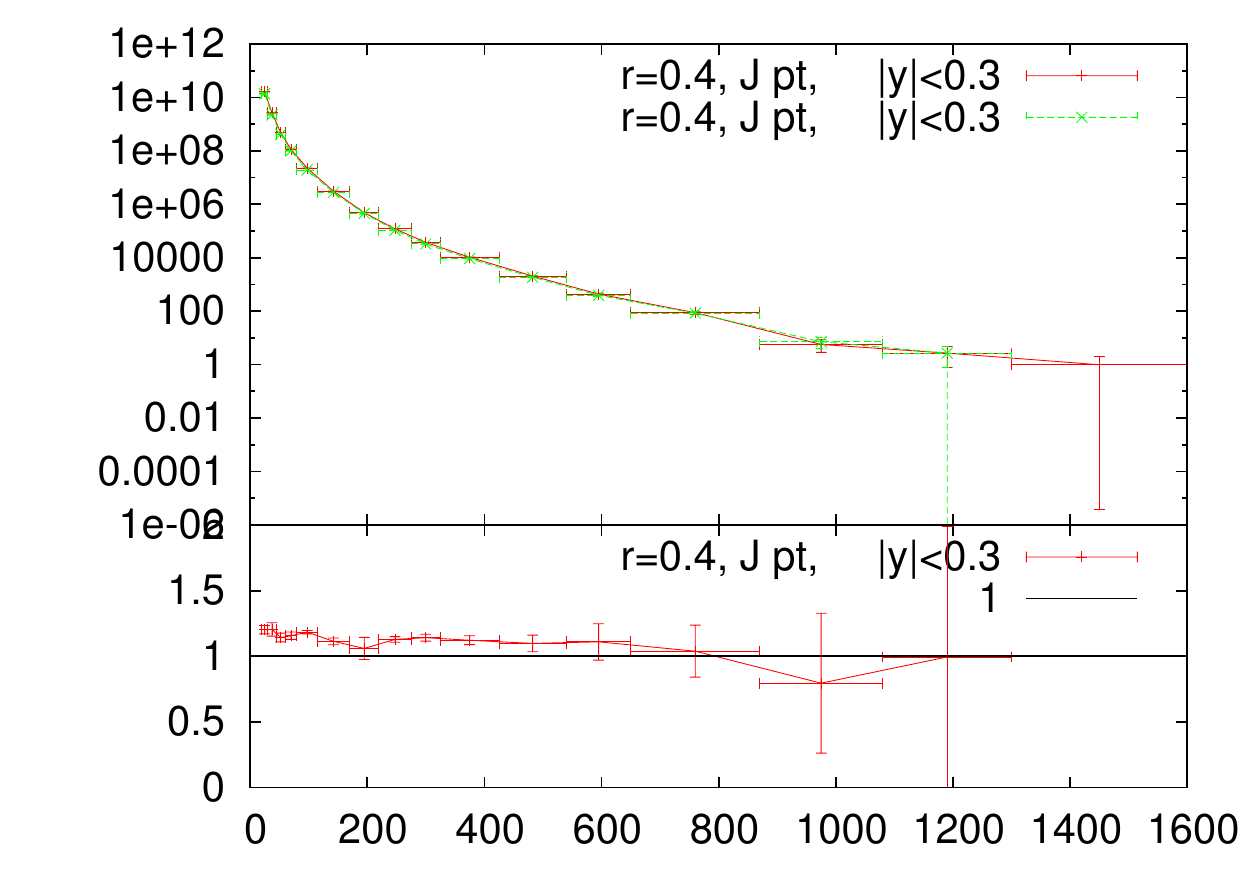,width=0.48\textwidth}
\epsfig{file=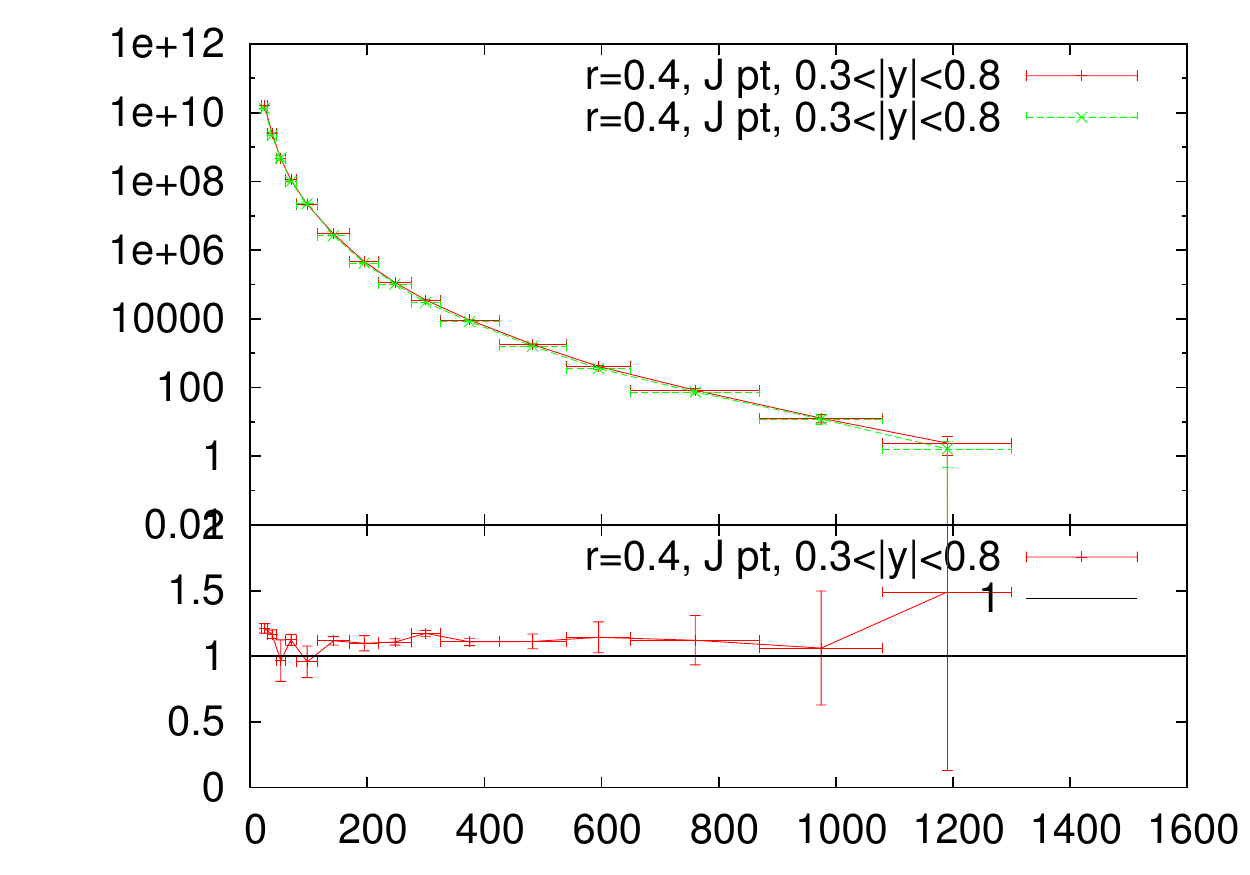,width=0.48\textwidth}
\epsfig{file=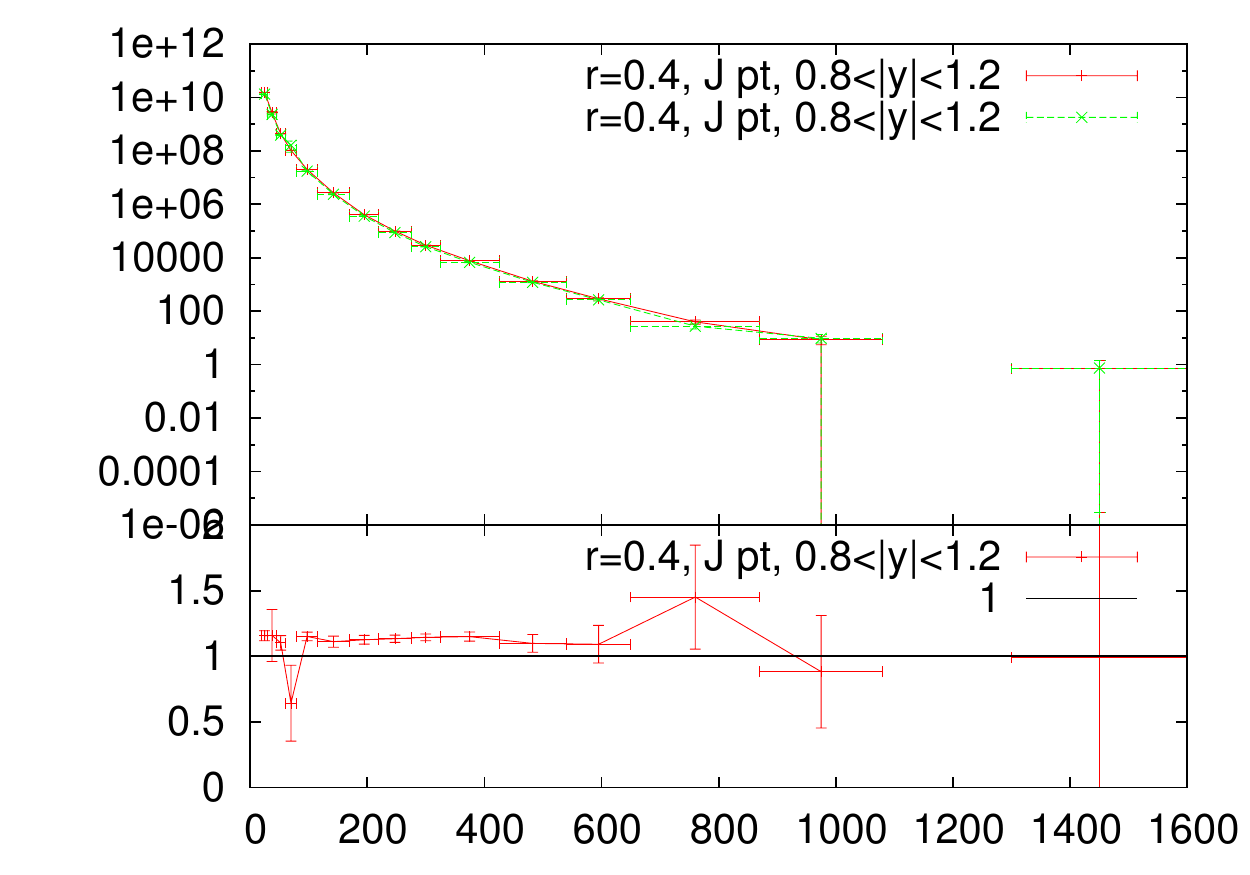,width=0.48\textwidth}
\epsfig{file=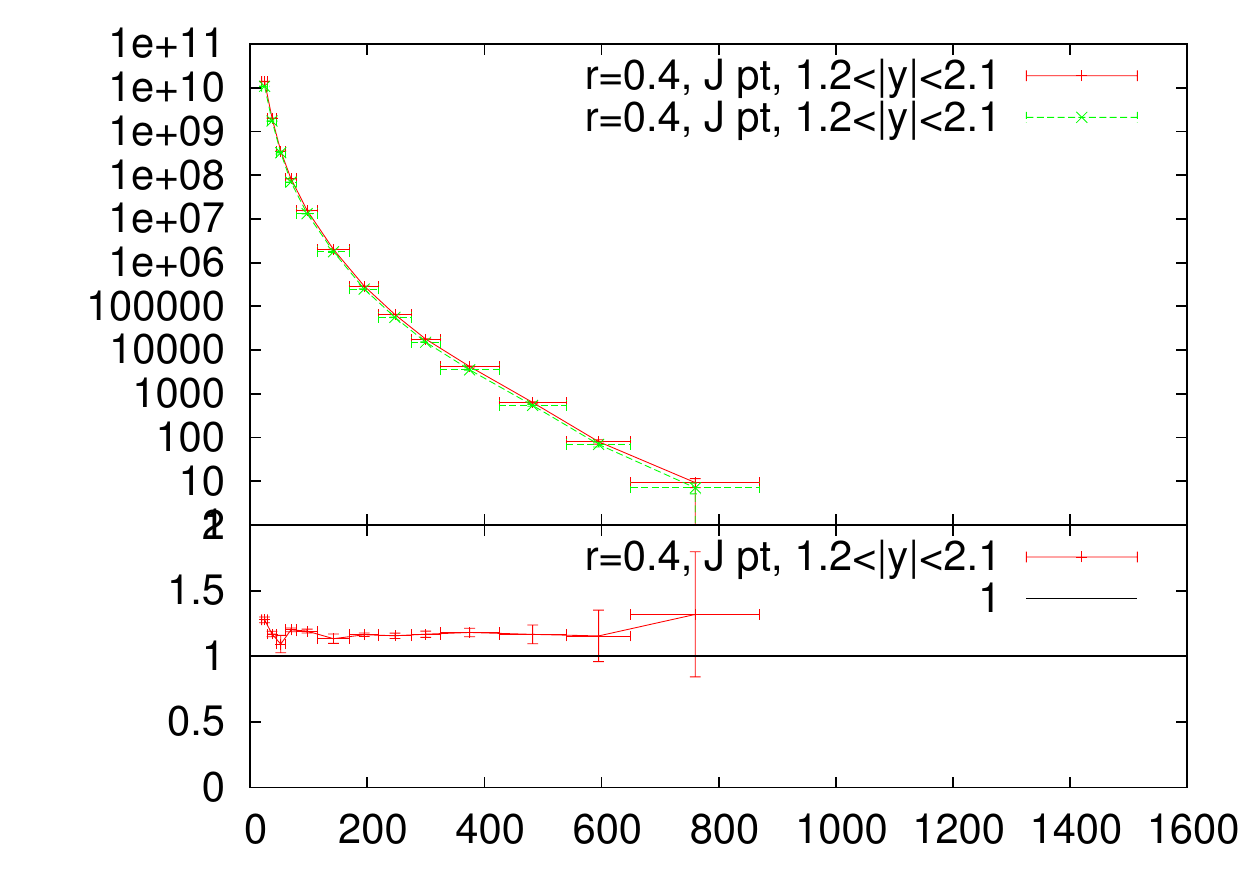,width=0.48\textwidth}
\epsfig{file=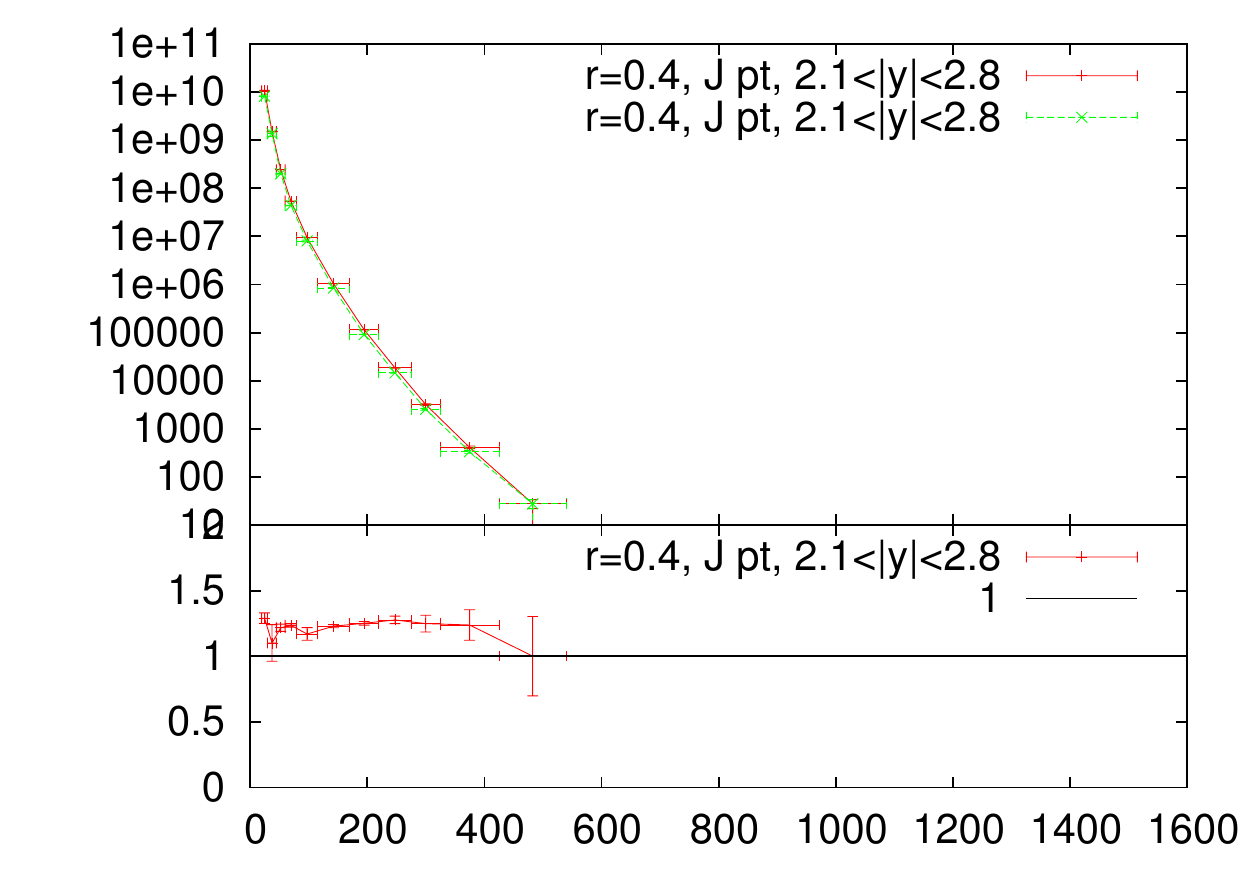,width=0.48\textwidth}
\epsfig{file=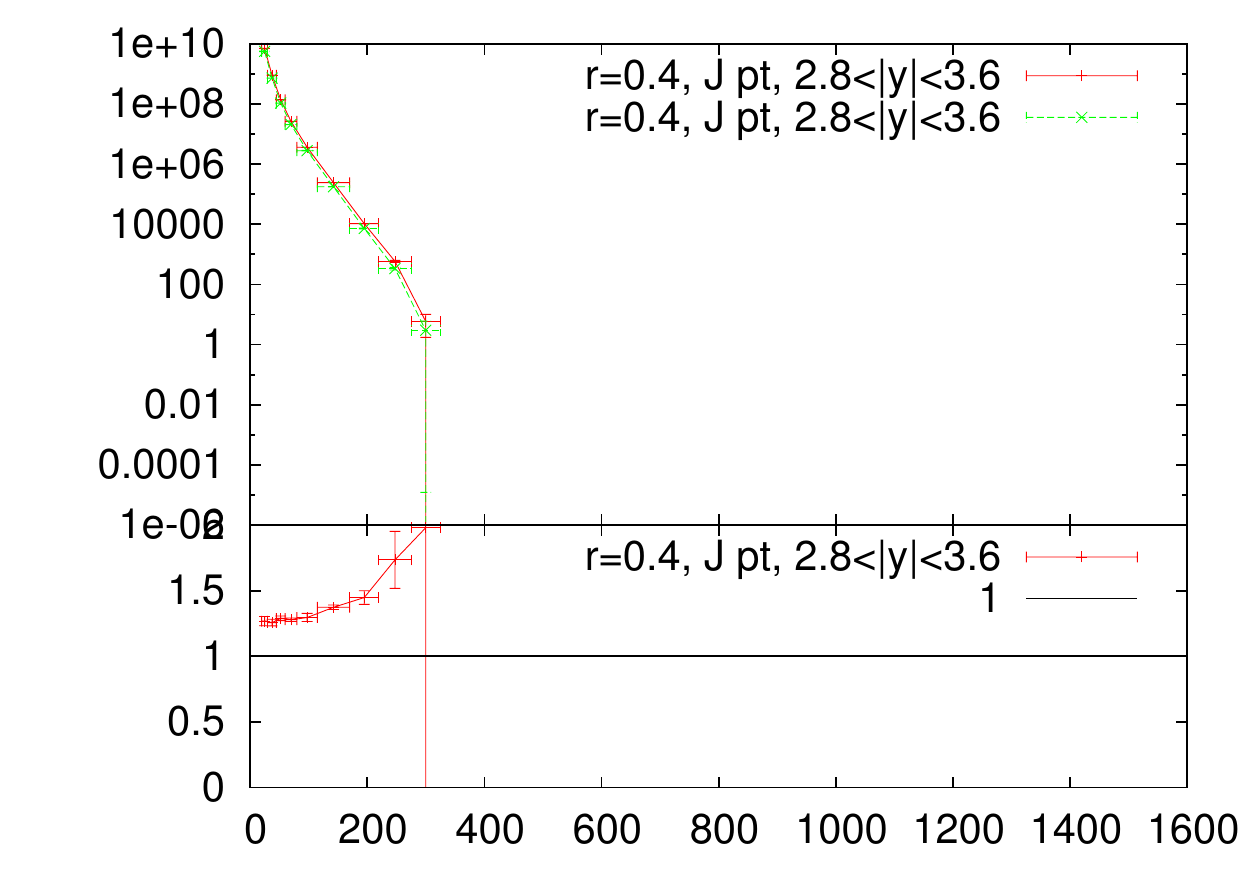,width=0.48\textwidth}
\end{center}
\caption{Comparison of the inclusive jet $\pt$ distribution in sample D1, at
  the Les Houches level (red), and after shower (green).}
\label{fig:pt3}
\end{figure}
In figs.~\ref{fig:pt2} and~\ref{fig:pt3} we compare the Les Houches versus
the showered results in the D0 and D1 samples.  We see that spikes are
present, but are much less important in the D1 sample.

\begin{figure}[htb]
\begin{center}
\epsfig{file=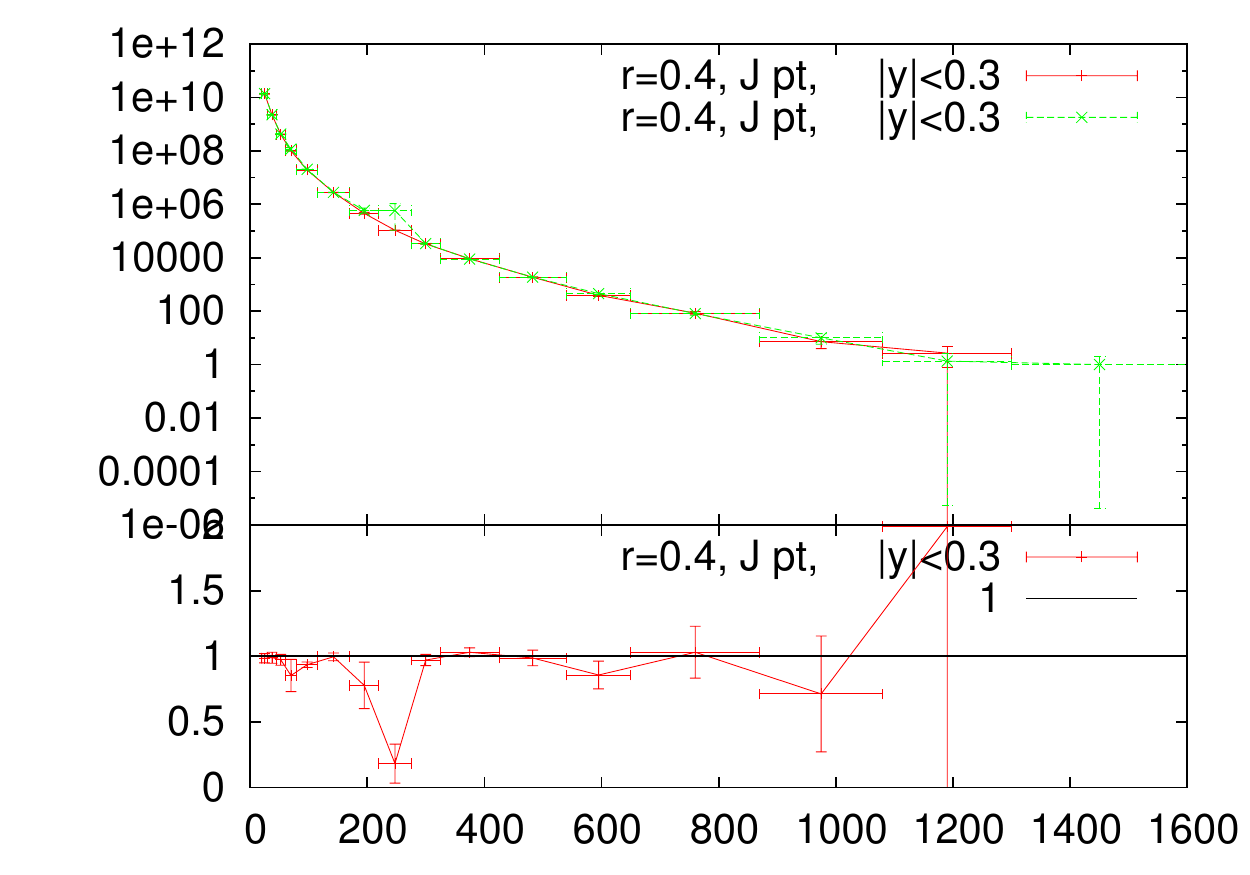,width=0.48\textwidth}
\epsfig{file=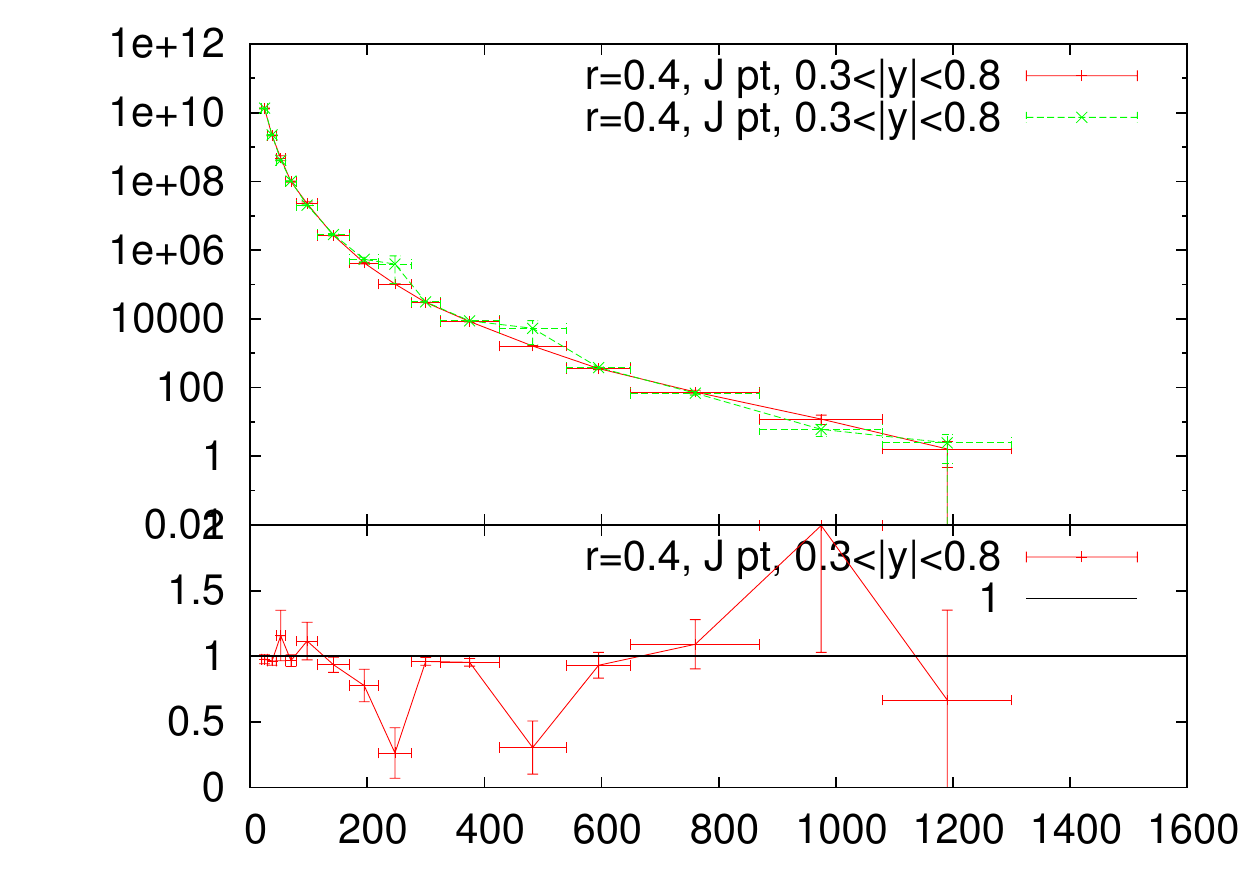,width=0.48\textwidth}
\epsfig{file=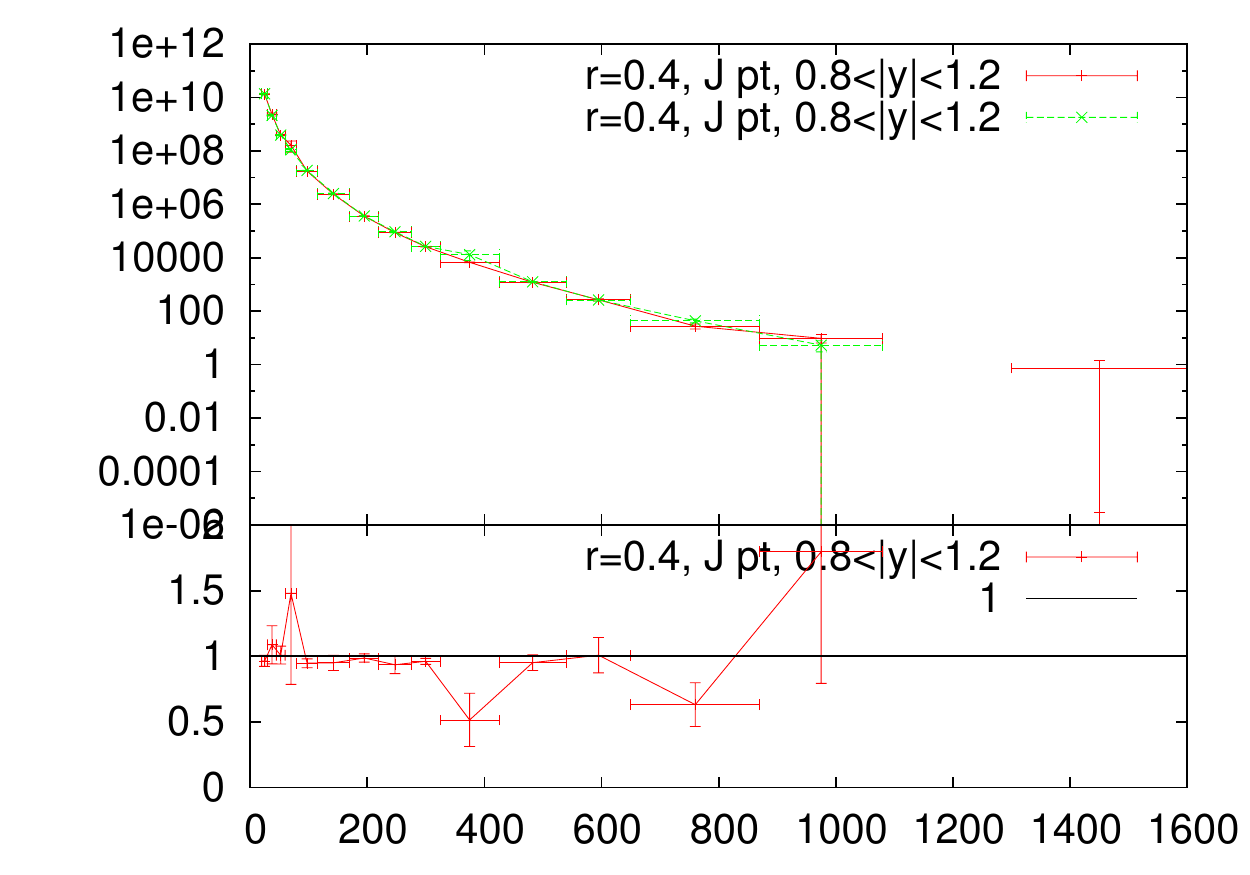,width=0.48\textwidth}
\epsfig{file=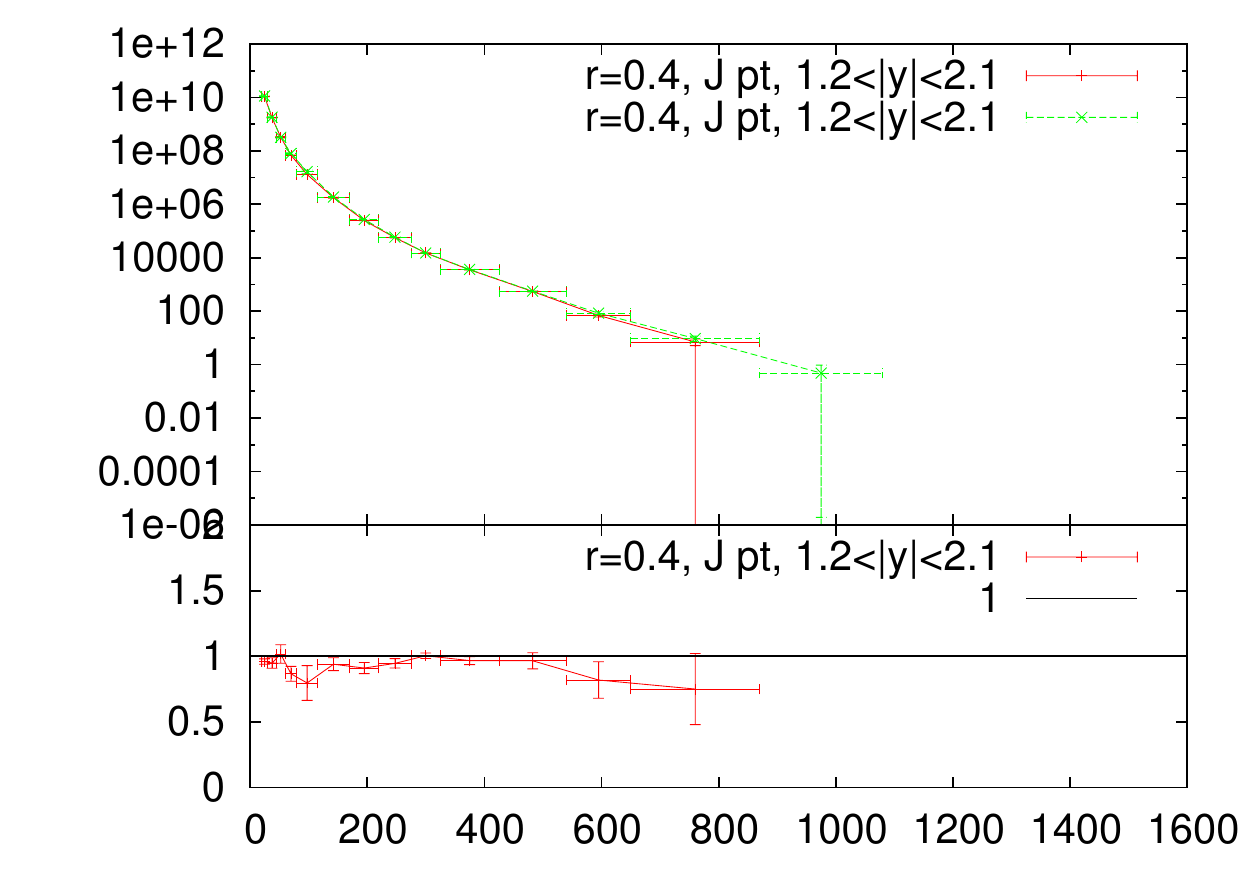,width=0.48\textwidth}
\epsfig{file=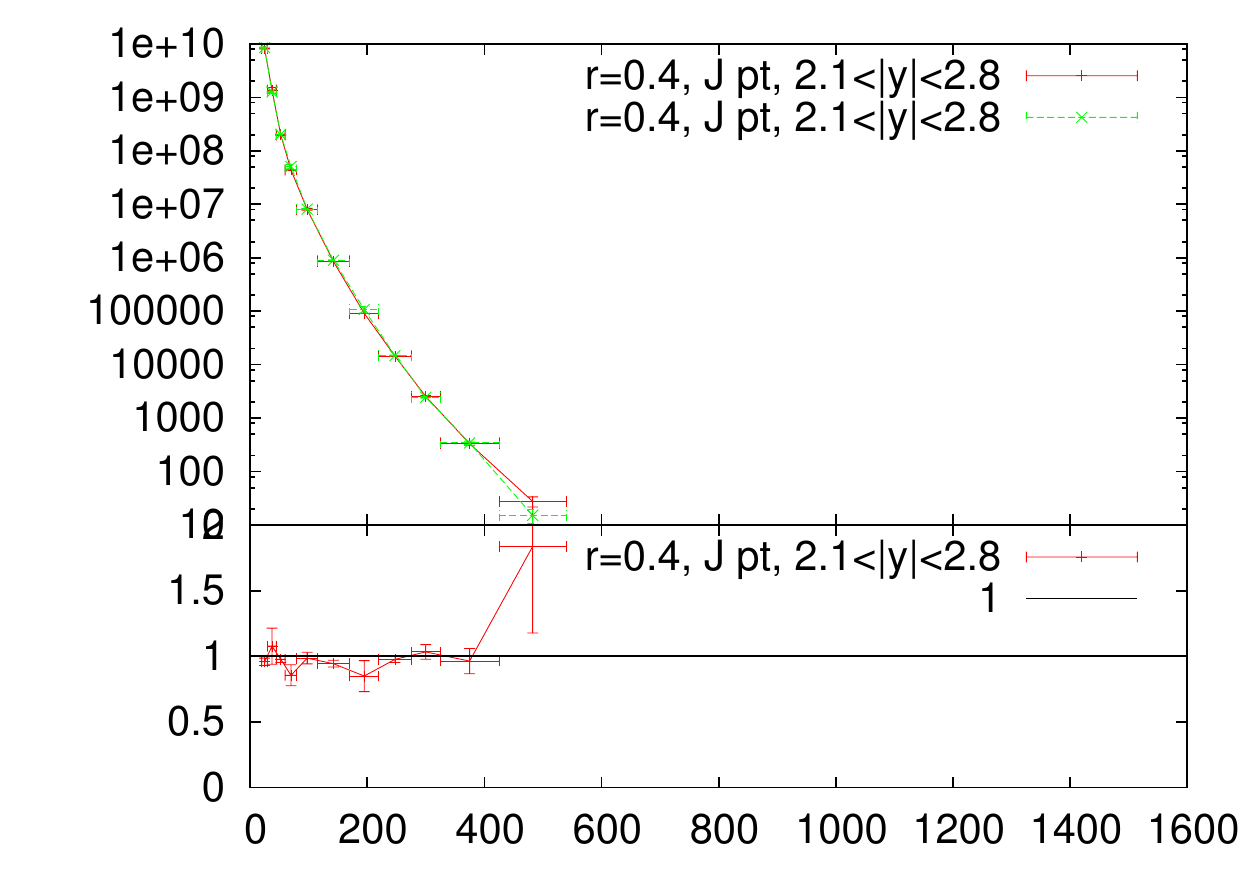,width=0.48\textwidth}
\epsfig{file=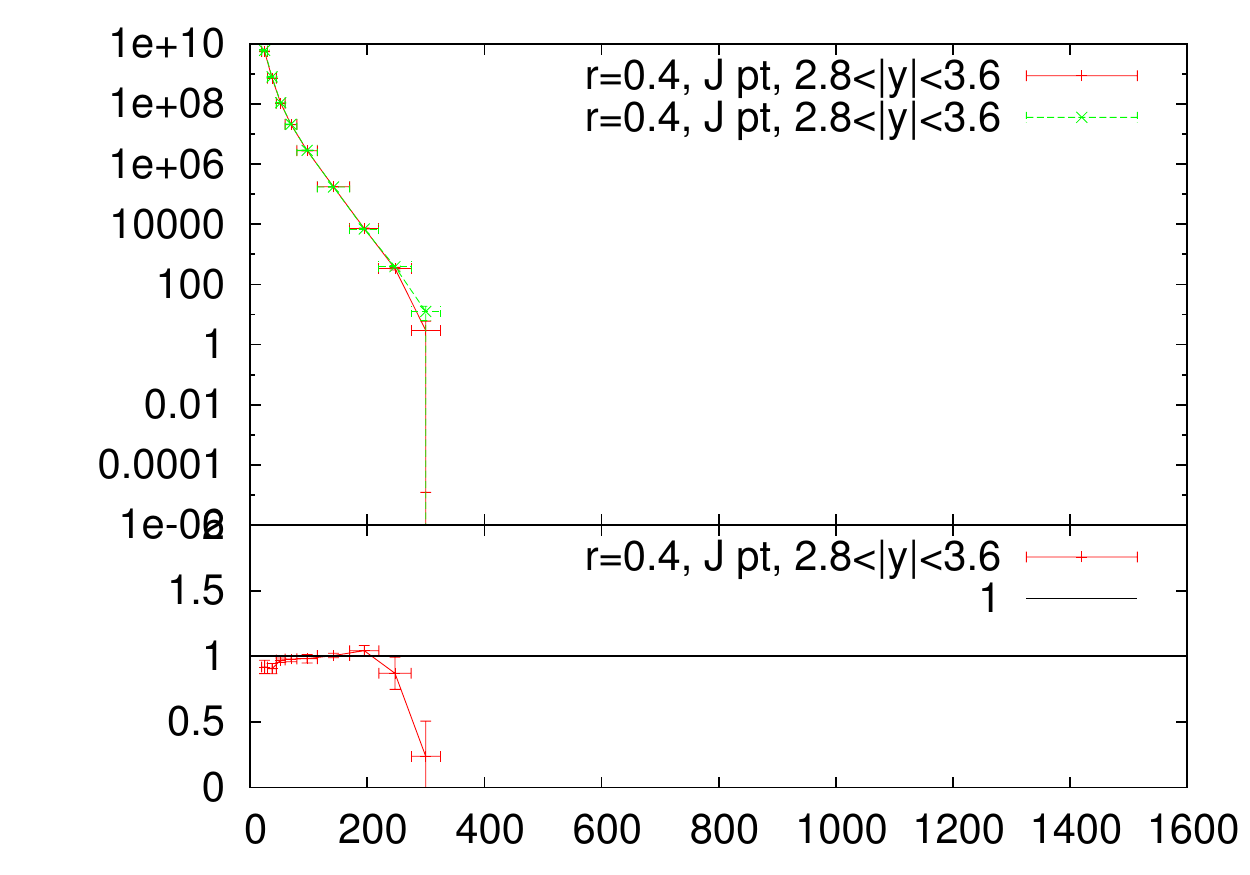,width=0.48\textwidth}
\end{center}
\caption{Comparison of the inclusive jet $\pt$ distribution from the showered
  sample D1 (red) and D0 (green).}
\label{fig:pt4}
\end{figure}
The trend in the contribution of shower effects seems to be similar in the D0
and D1 samples. It is thus interesting to compare the two samples directly,
to see if drastic changes of the final distributions are present.  This
comparison is shown in fig.~\ref{fig:pt4}.  It is clear from the comparison
that, aside from the large spikes, no significant differences, above a few
percent, are visible in the two samples.

\begin{figure}[htb]
\begin{center}
\epsfig{file=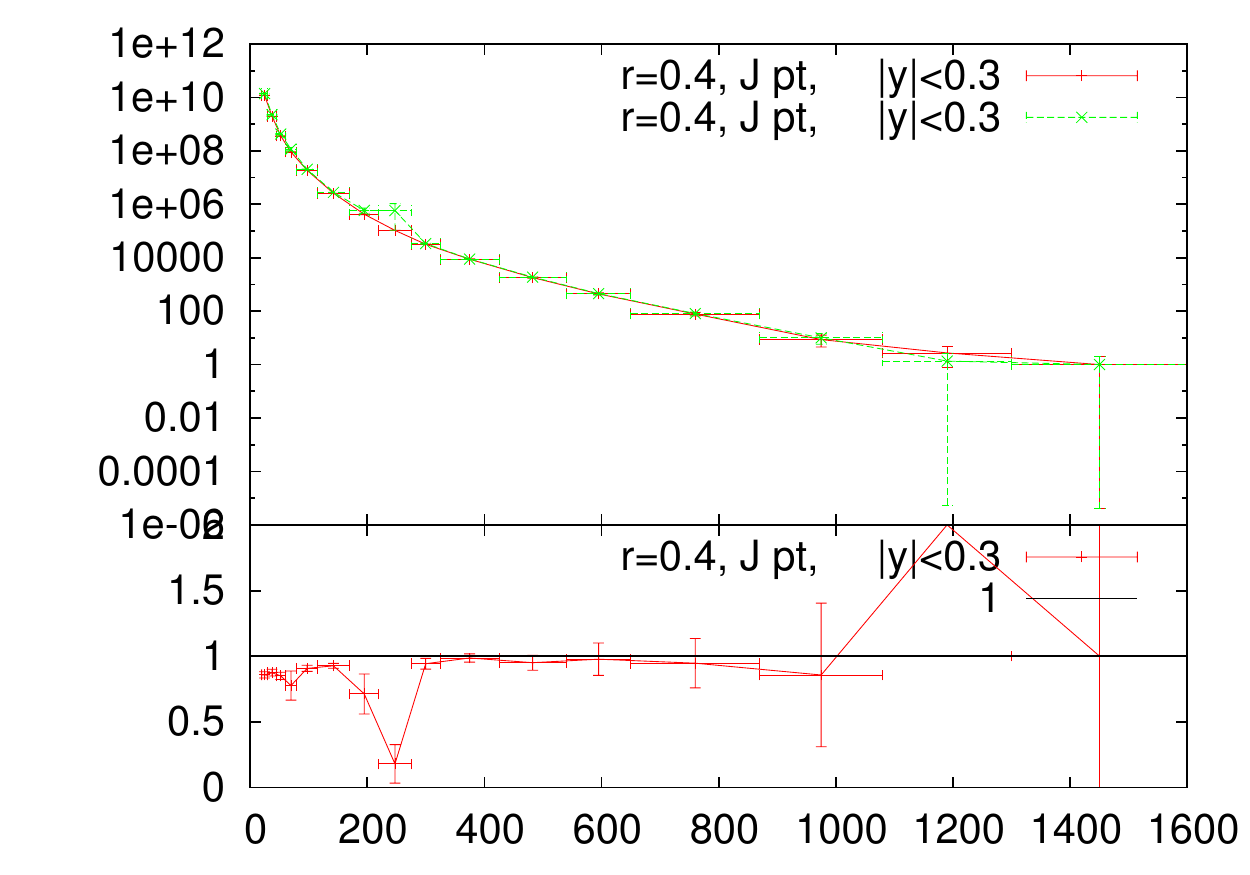,width=0.48\textwidth}
\epsfig{file=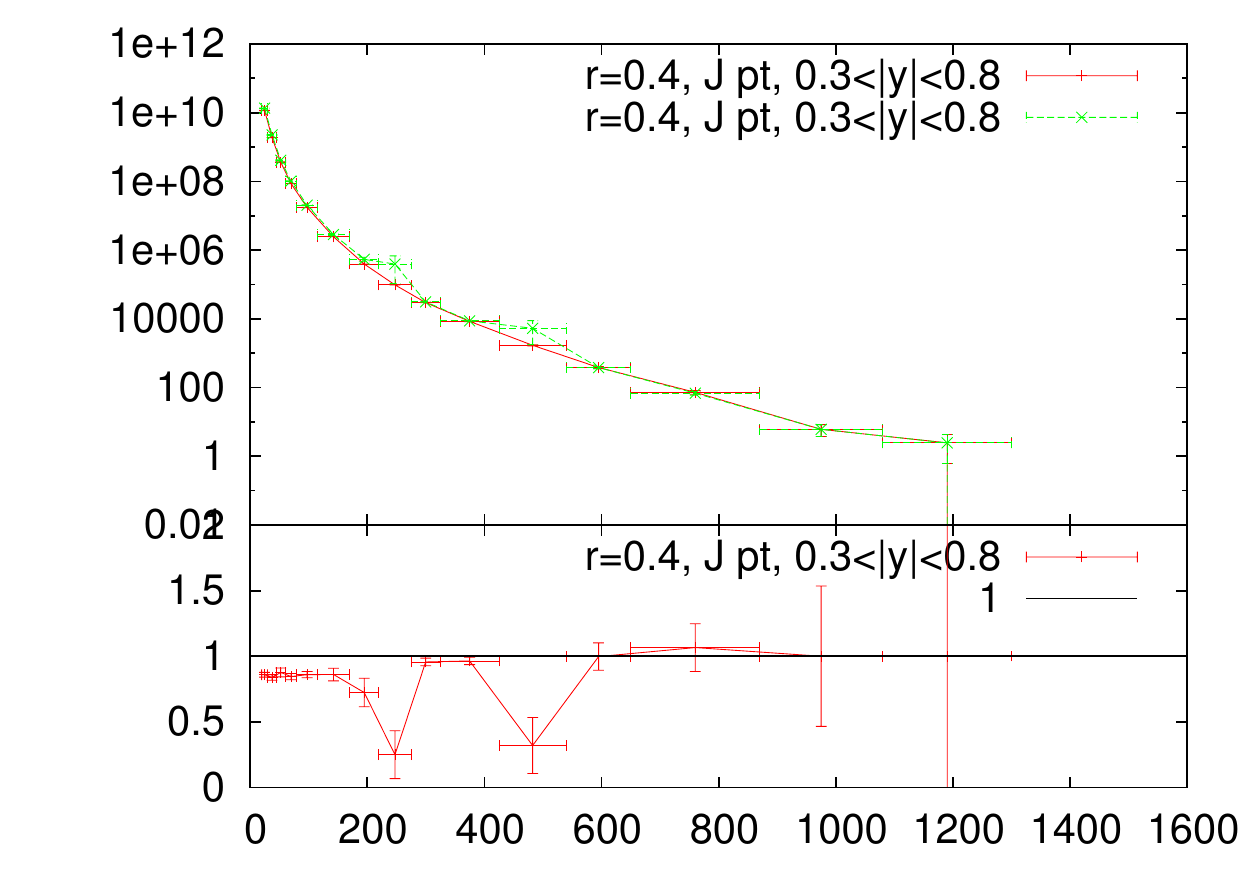,width=0.48\textwidth}
\epsfig{file=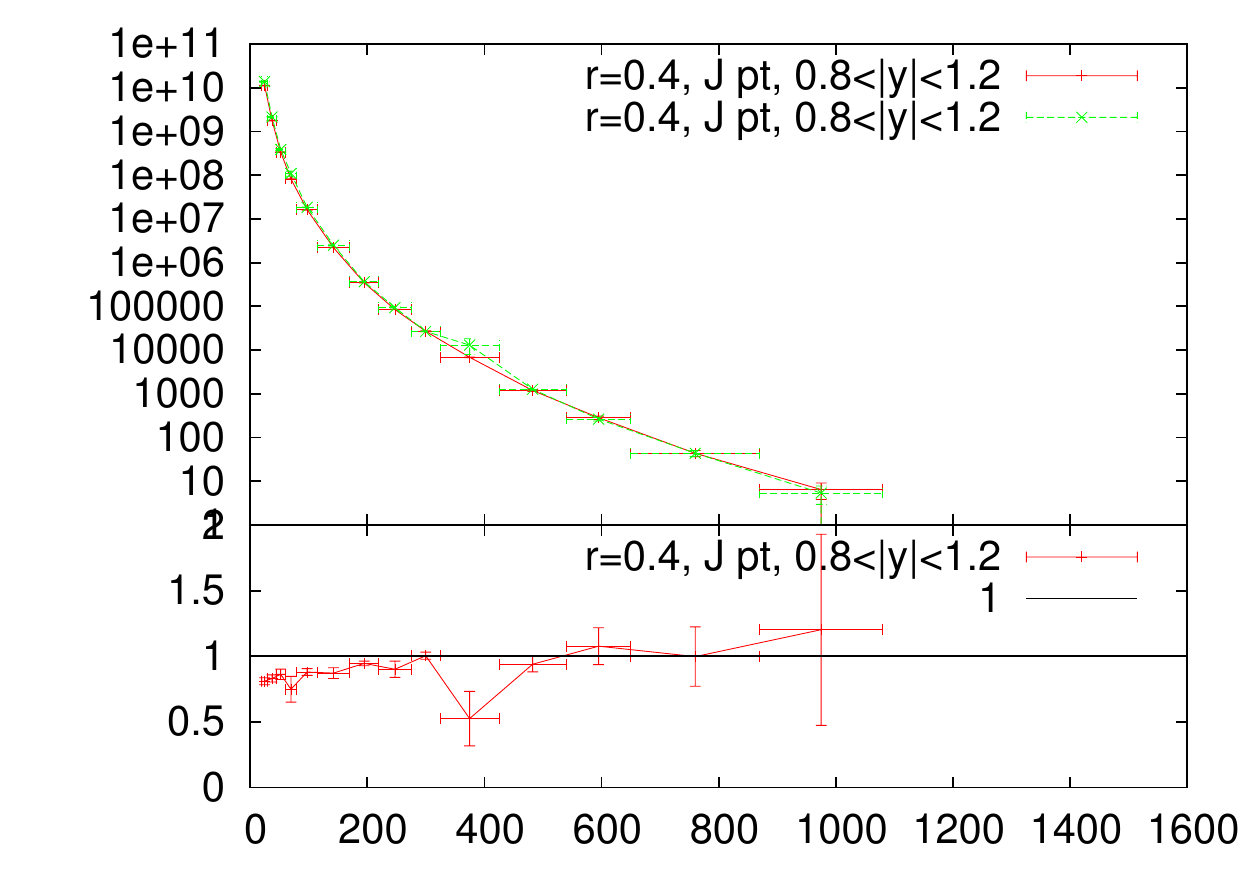,width=0.48\textwidth}
\epsfig{file=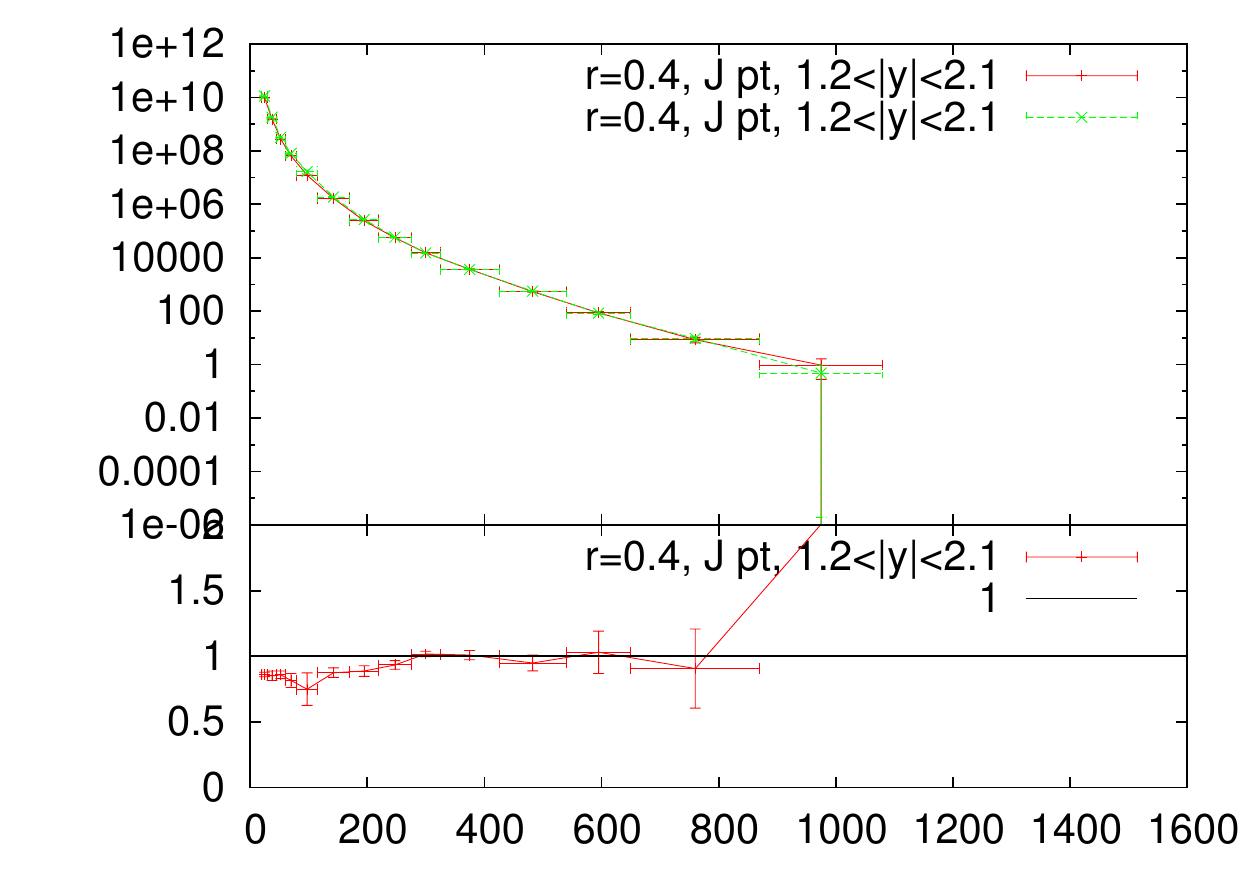,width=0.48\textwidth}
\epsfig{file=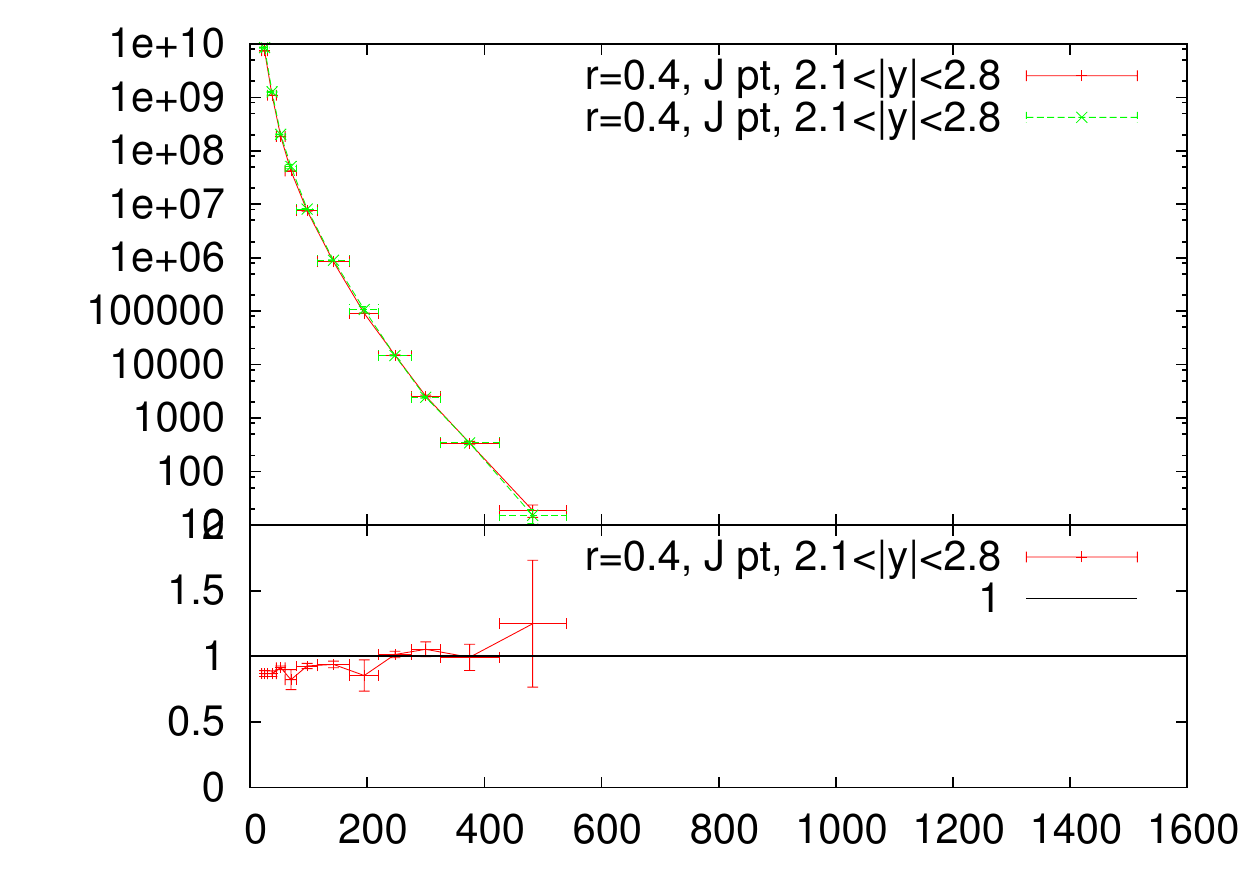,width=0.48\textwidth}
\epsfig{file=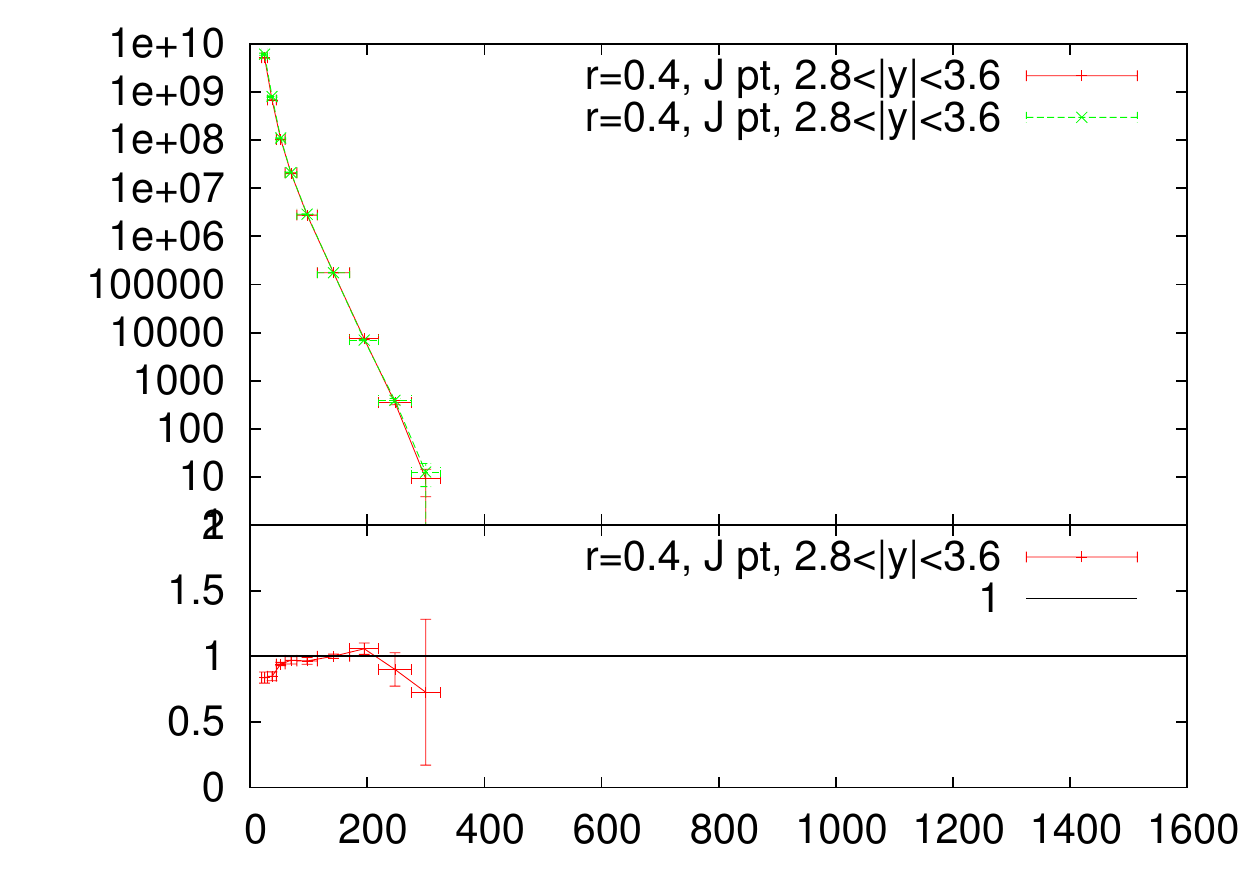,width=0.48\textwidth}
\end{center}
\caption{Comparison of the inclusive jet $\pt$ distribution
from the showered D0 sample. We compare the output obtained
using the RS choice for {\tt scalup} (red) versus the standard one
(green).}
\label{fig:pt5}
\end{figure}
\begin{figure}[htb]
\begin{center}
\epsfig{file=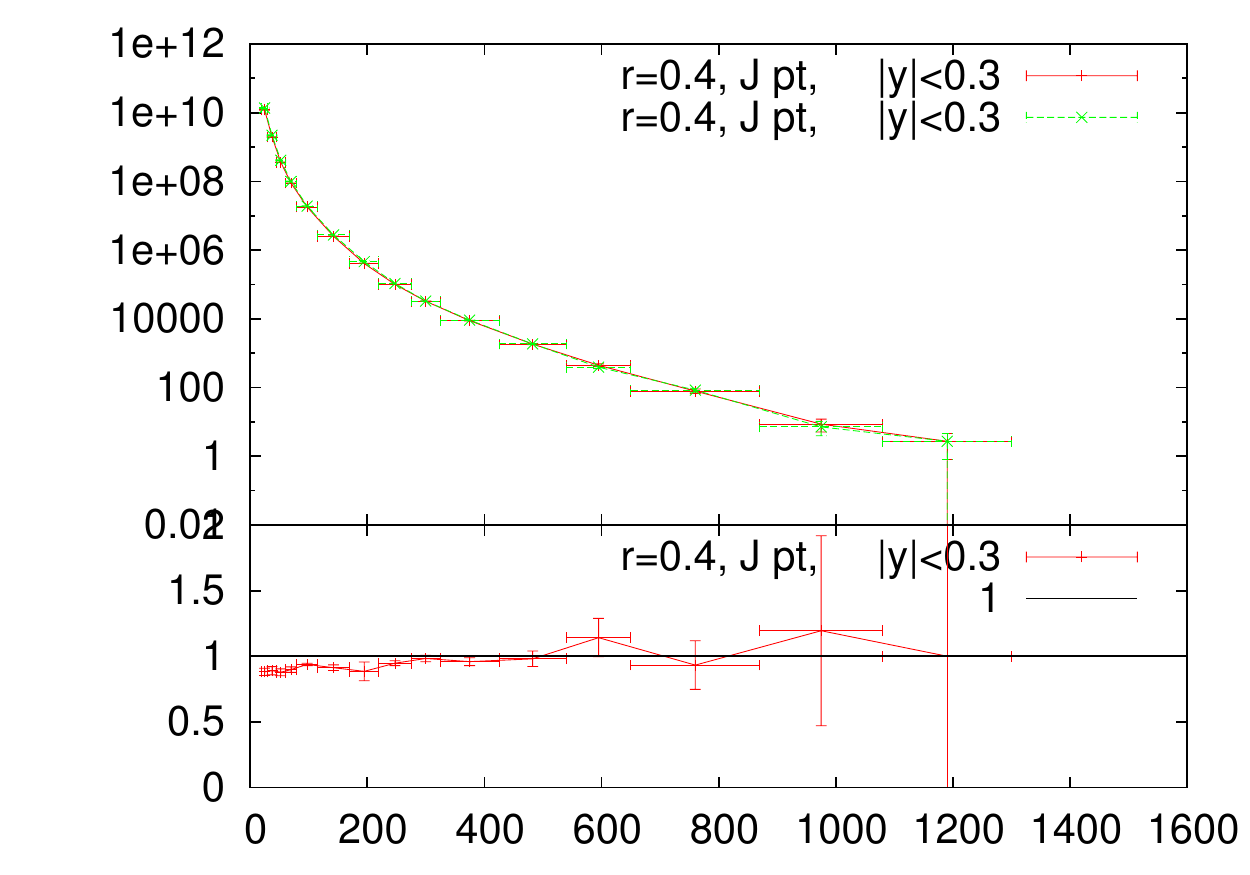,width=0.48\textwidth}
\epsfig{file=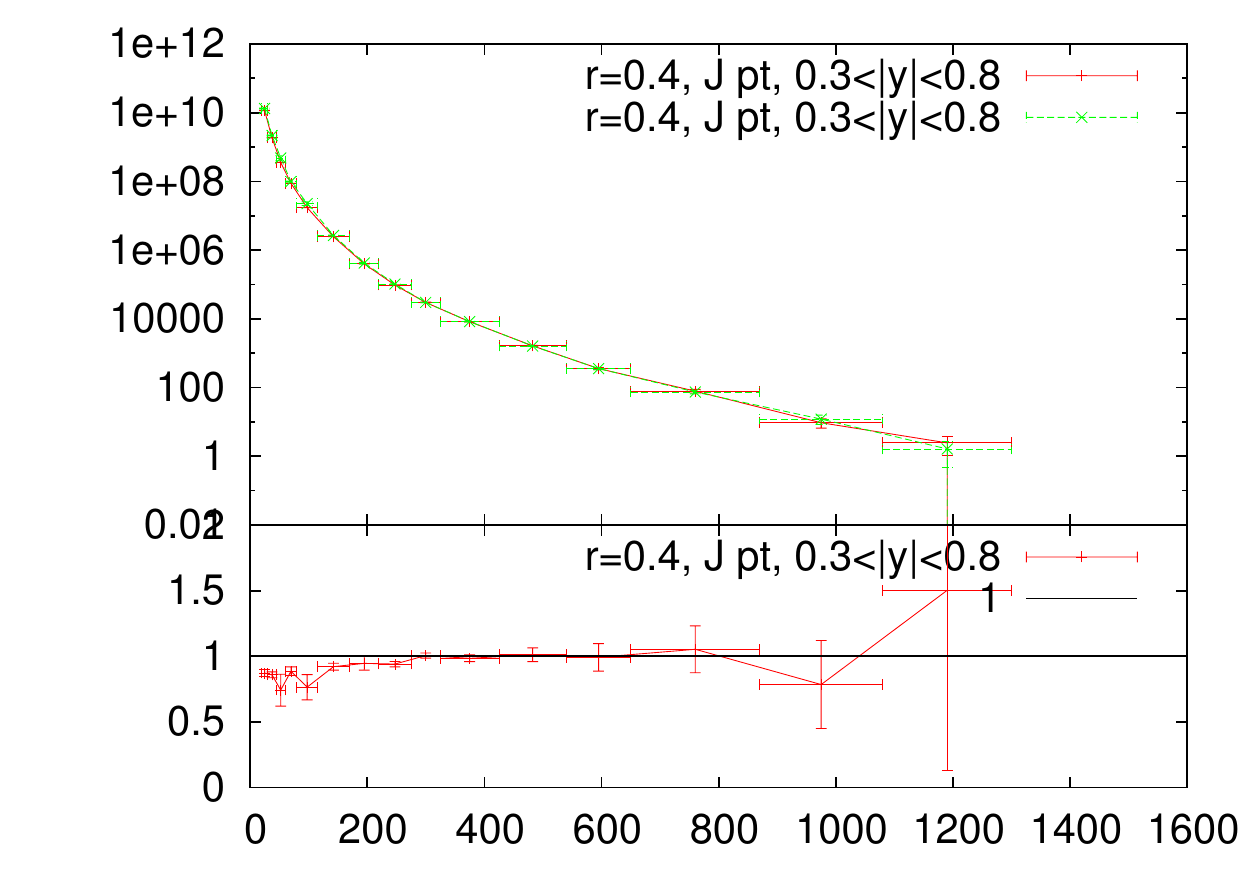,width=0.48\textwidth}
\epsfig{file=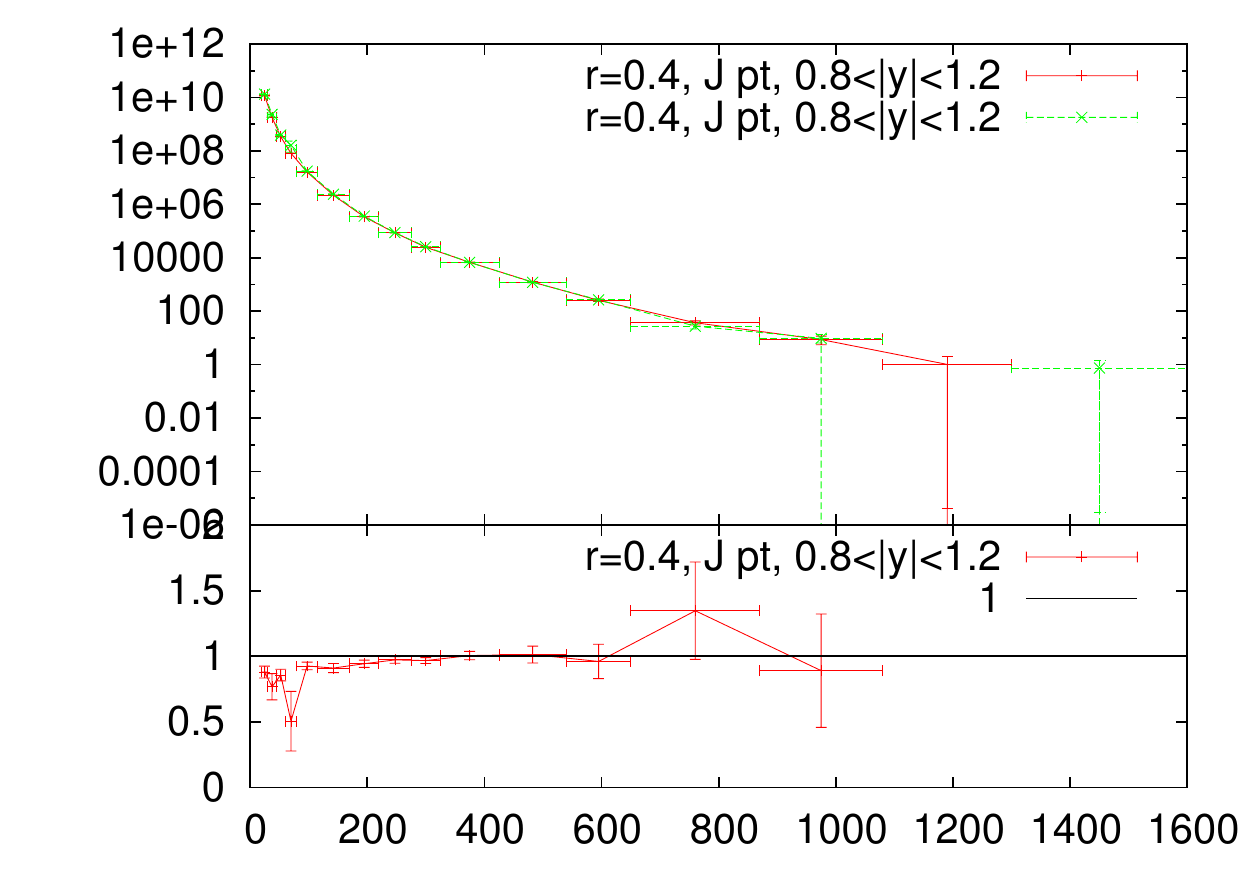,width=0.48\textwidth}
\epsfig{file=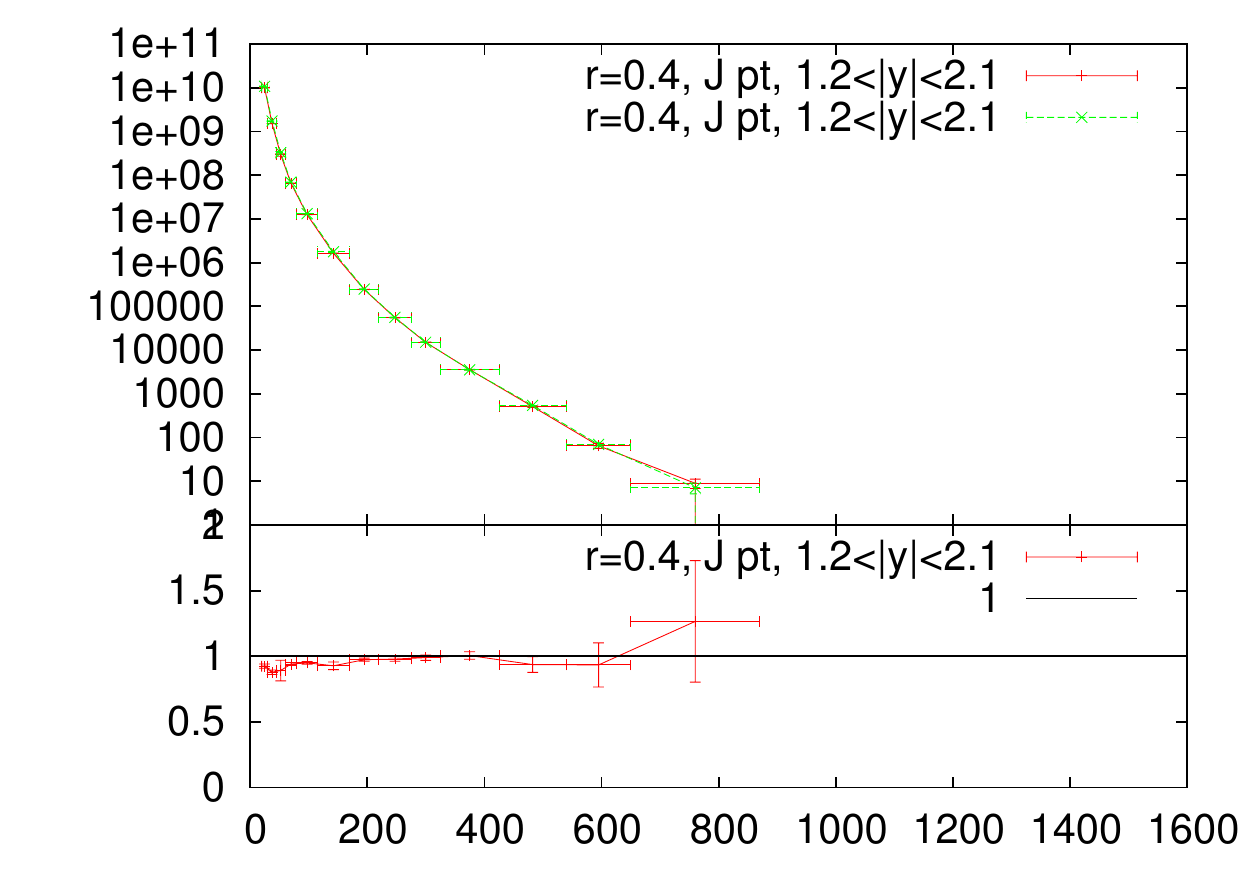,width=0.48\textwidth}
\epsfig{file=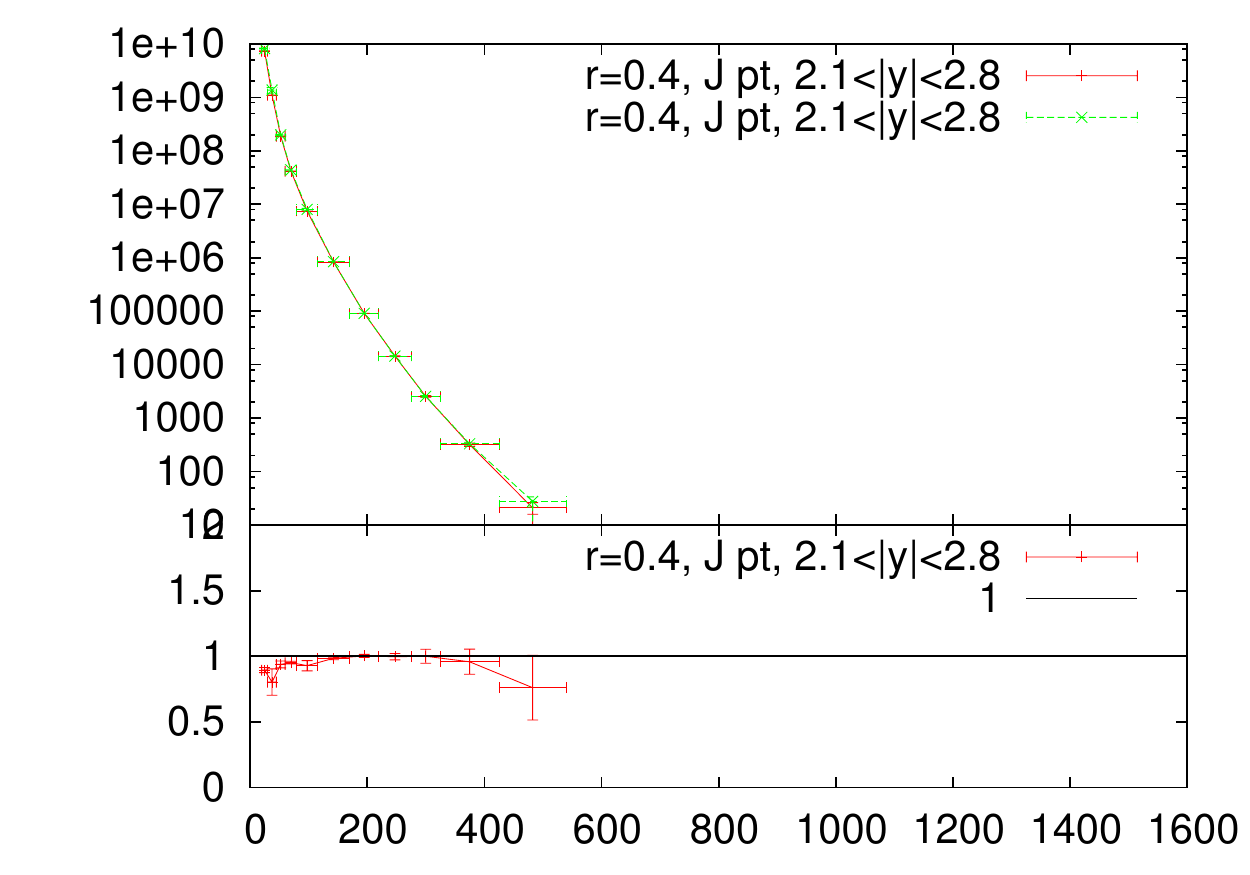,width=0.48\textwidth}
\epsfig{file=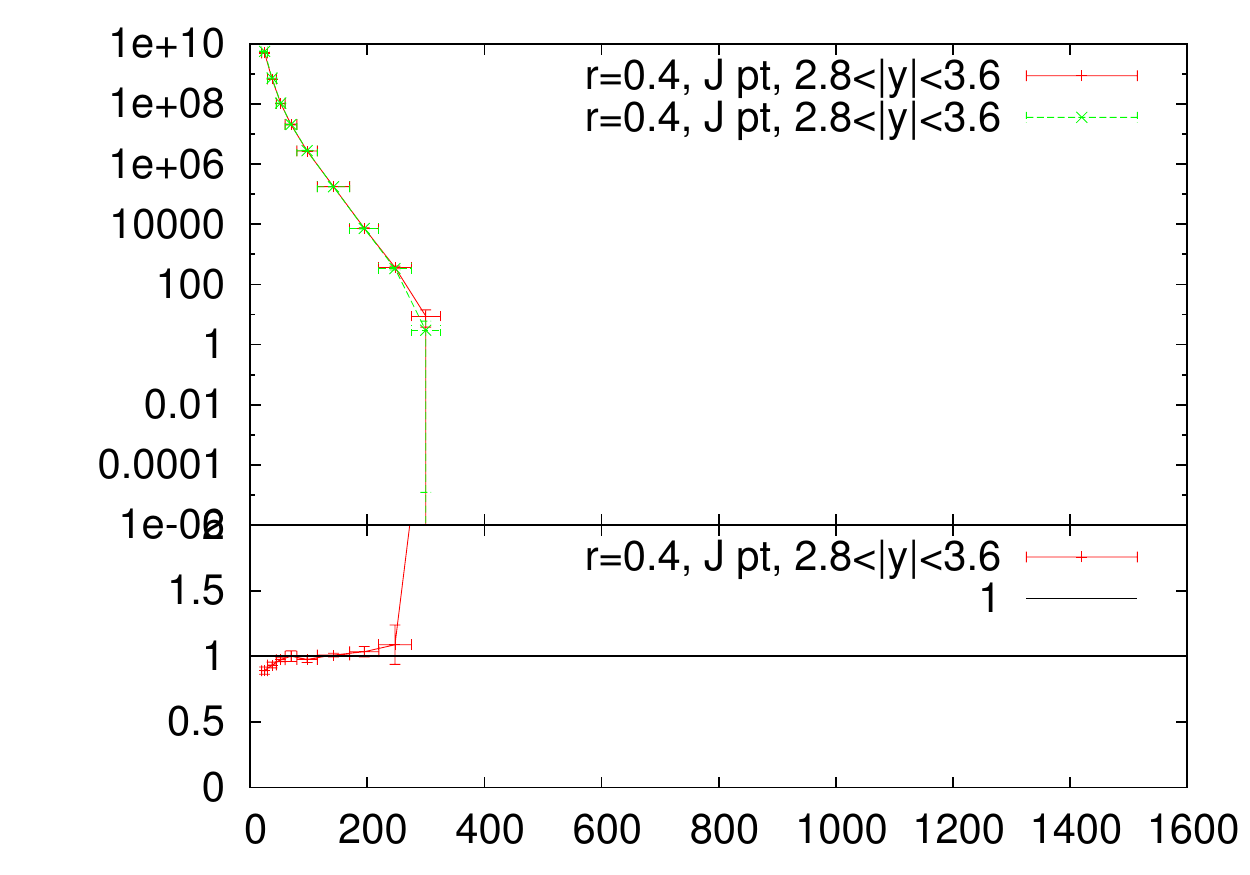,width=0.48\textwidth}
\end{center}
\caption{Comparison of the inclusive jet $\pt$ distribution from the showered
  D1 sample. We compare the output obtained using the RS choice for {\tt
    scalup} (red) versus the standard one (green).}
\label{fig:pt6}
\end{figure}
We now consider the effect of changing the way that the upper
bound for radiation in the shower is computed.  We show in
figs.~\ref{fig:pt5} and~\ref{fig:pt6} the comparison of the result with the
the RS choice for {\tt scalup} versus the standard one, in sample D0
(fig.~\ref{fig:pt5}) and D1 (fig.~\ref{fig:pt6}).  We see that in both cases,
the use of the RS choice leads to some differences that can reach the 10\%{}
level for small transverse momenta.

\begin{figure}[htb]
\begin{center}
\epsfig{file=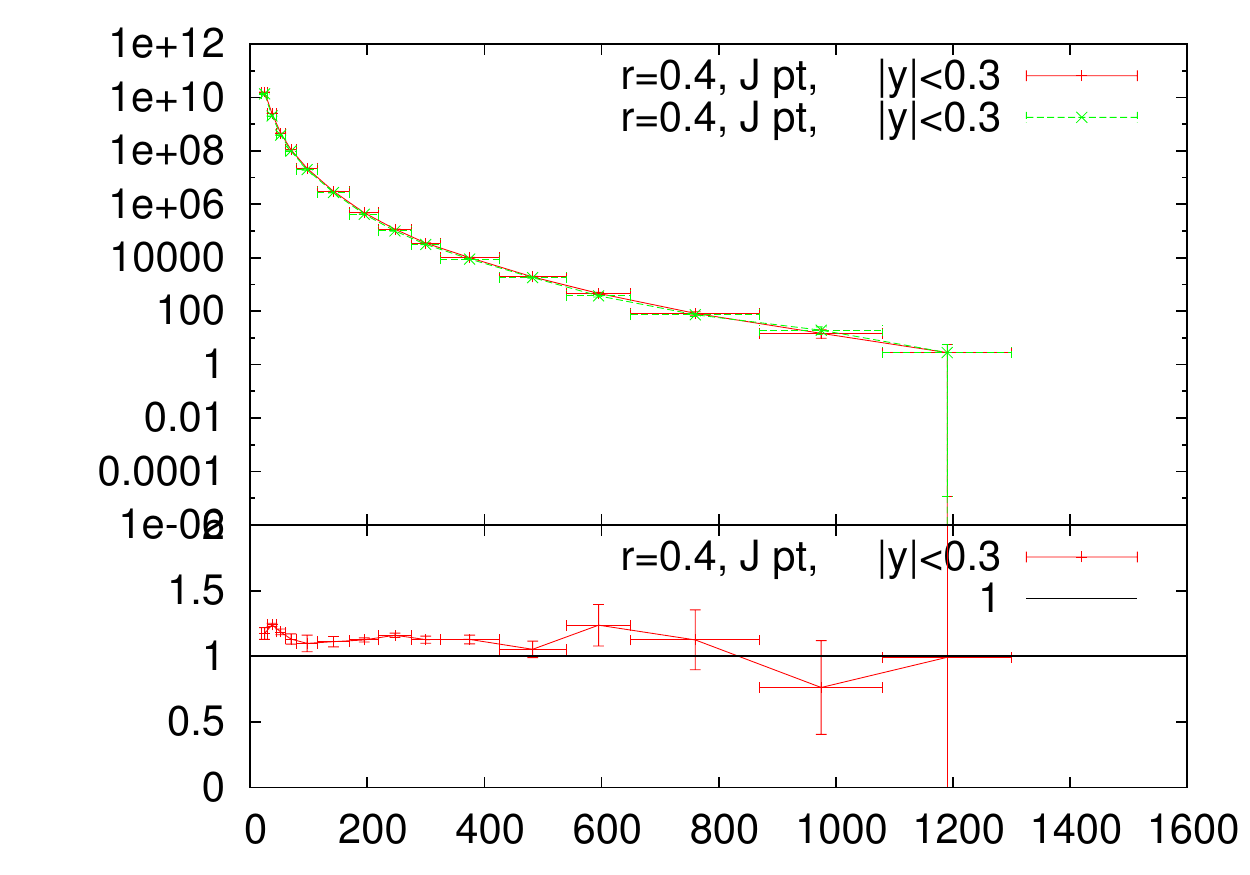,width=0.48\textwidth}
\epsfig{file=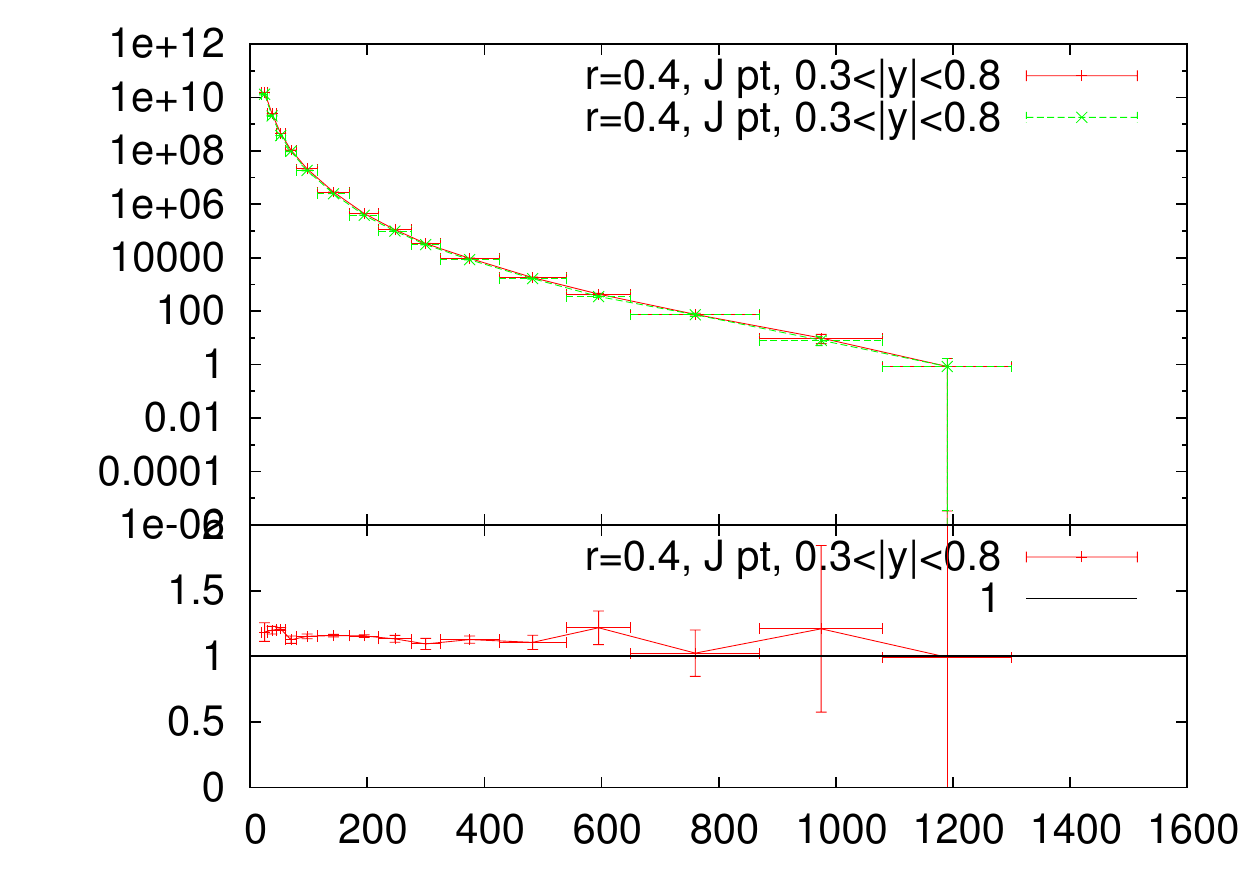,width=0.48\textwidth}
\epsfig{file=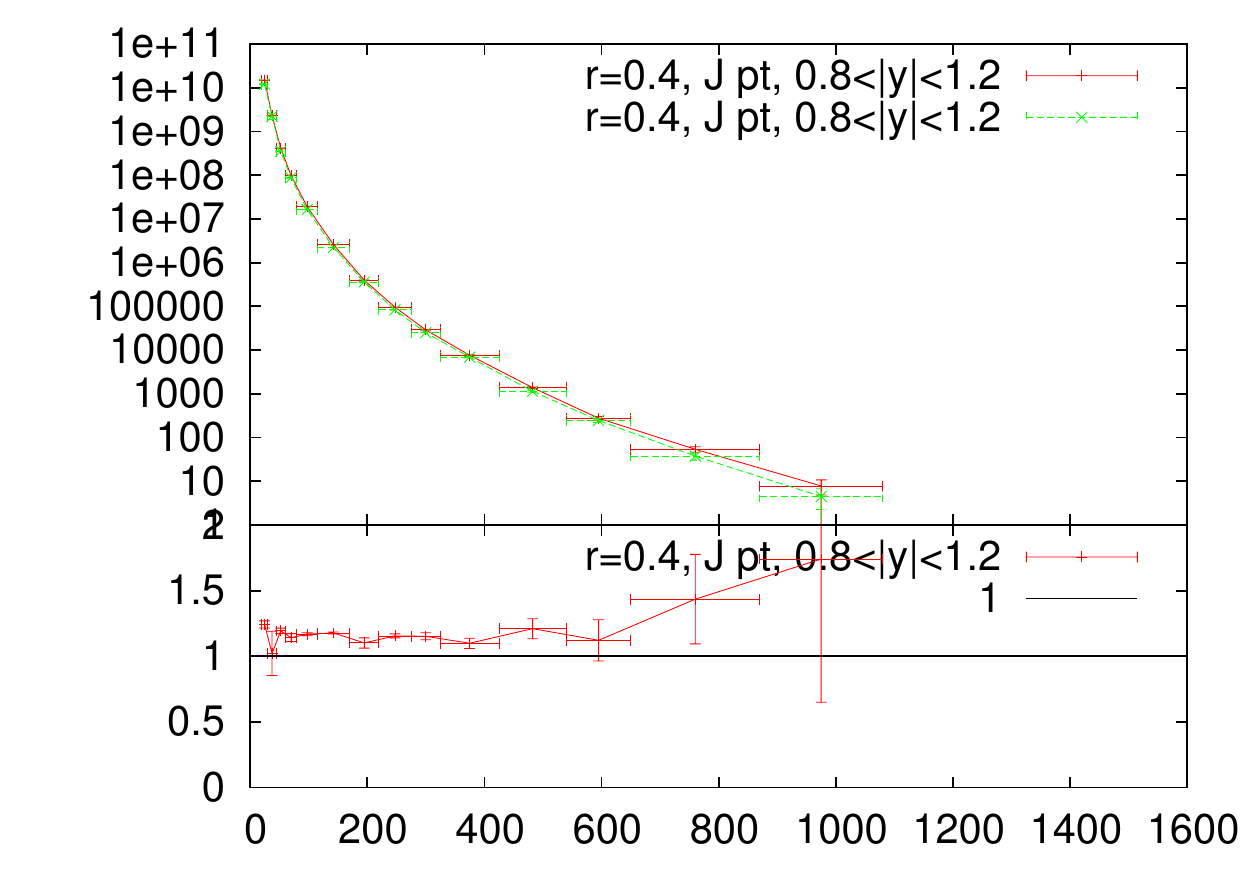,width=0.48\textwidth}
\epsfig{file=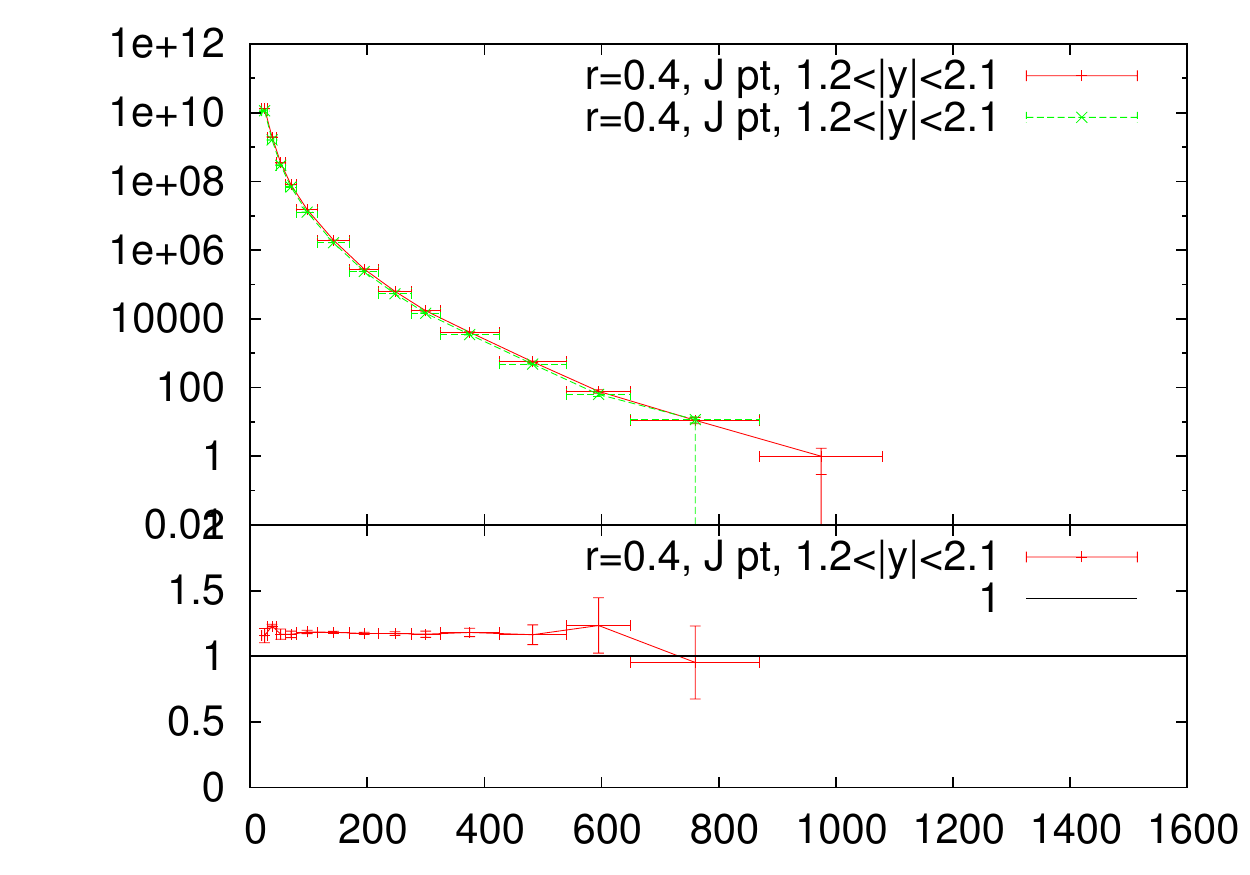,width=0.48\textwidth}
\epsfig{file=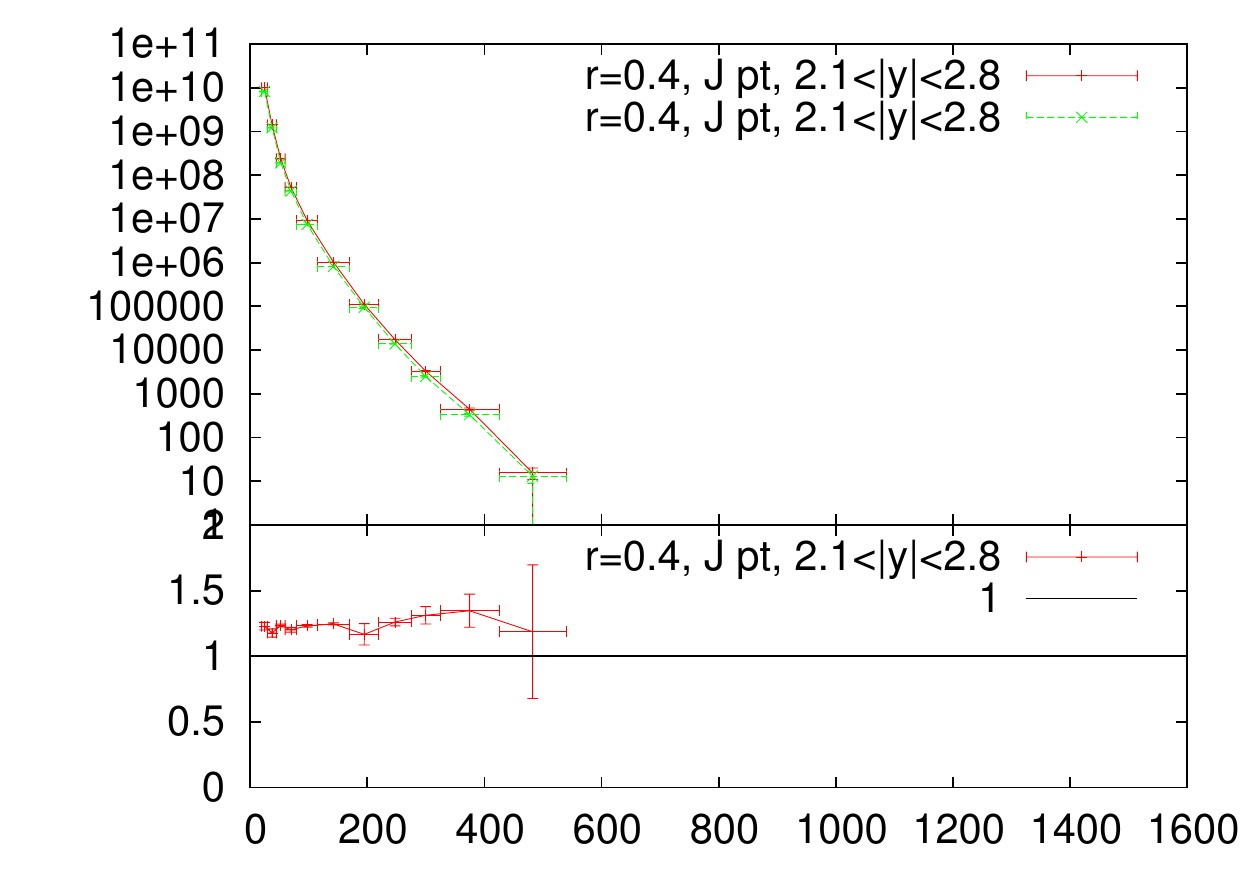,width=0.48\textwidth}
\epsfig{file=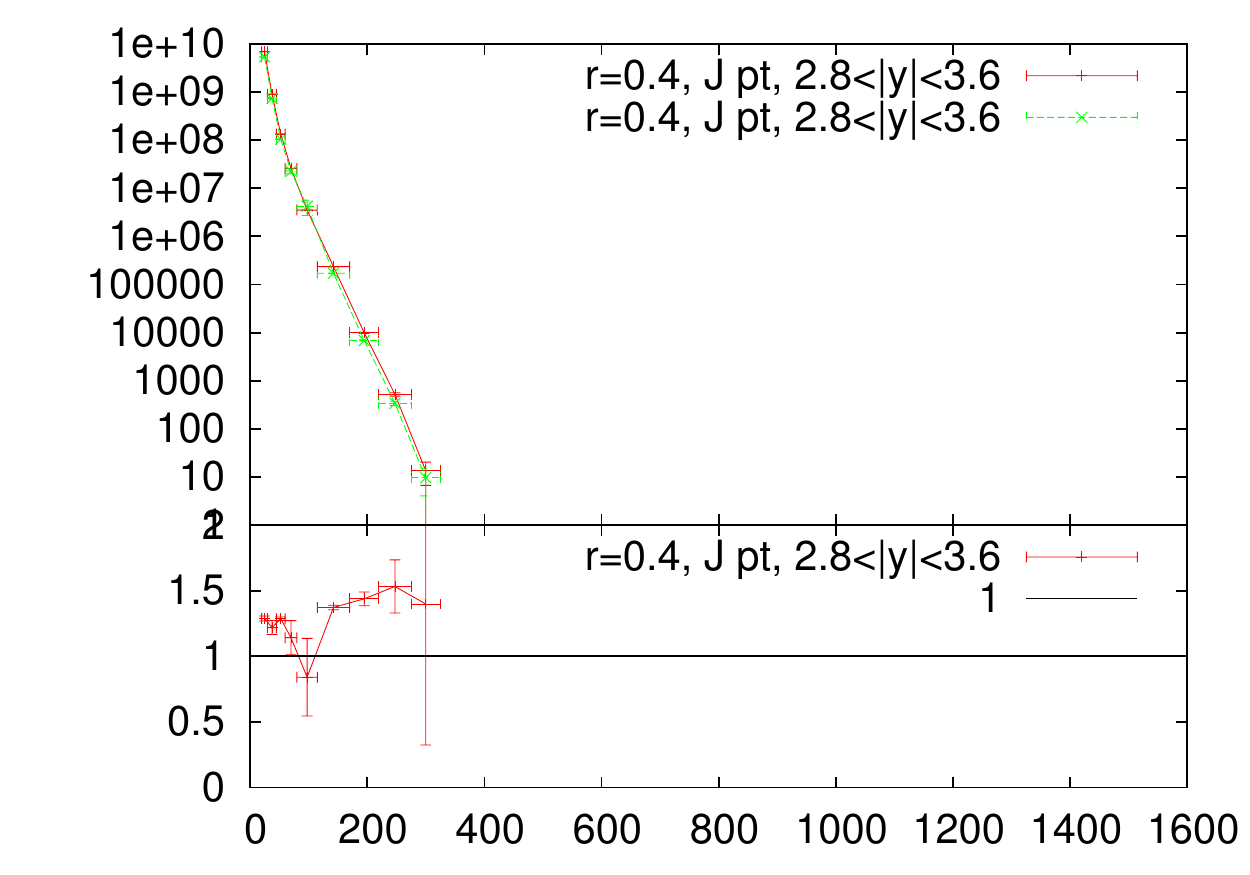,width=0.48\textwidth}
\end{center}
\caption{Comparison of the inclusive jet $\pt$ distribution, at the Les
  Houches level (red) and after shower (green), for the sample obtained with
  the {\tt doublefsr 1} option and with the new separation of regions.}
\label{fig:pt7}
\end{figure}

\begin{figure}[htb]
\begin{center}
\epsfig{file=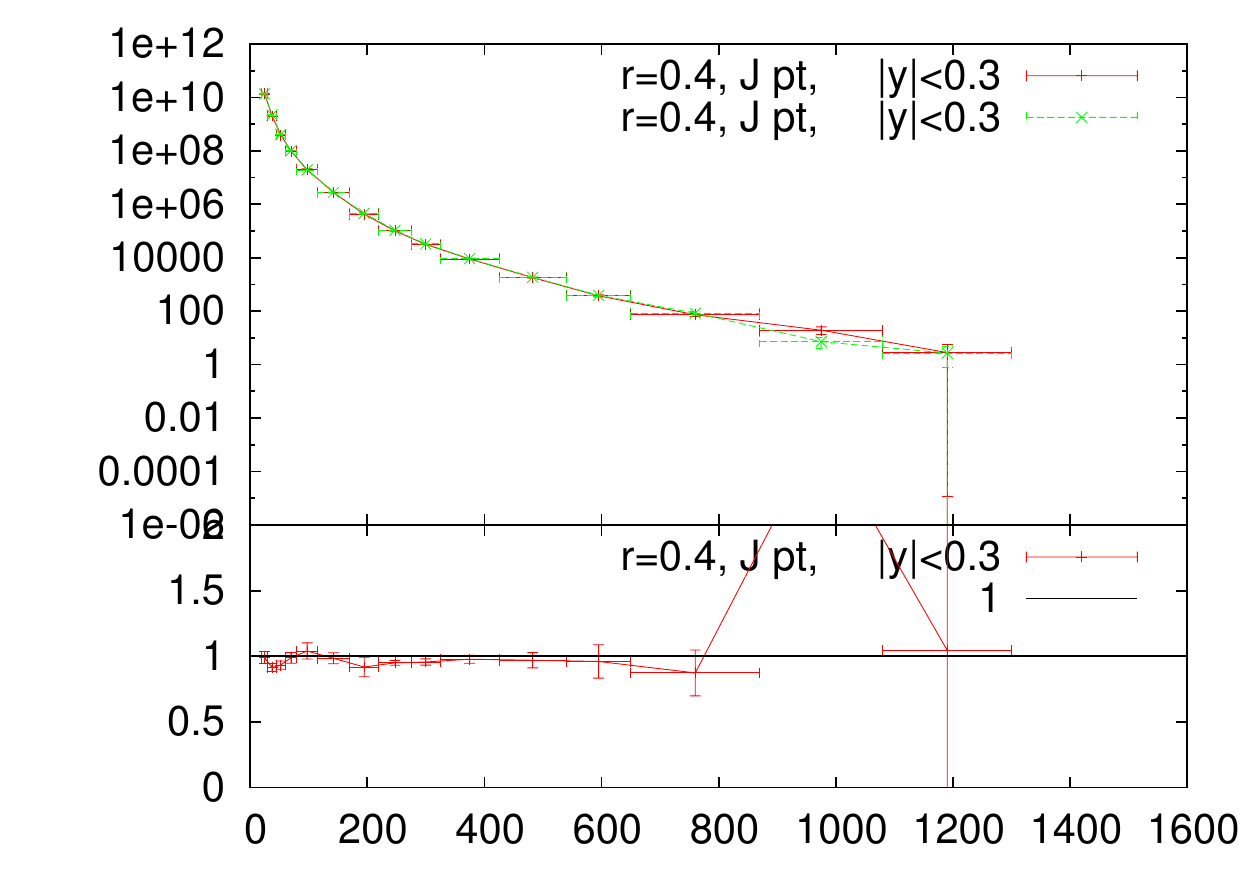,width=0.48\textwidth}
\epsfig{file=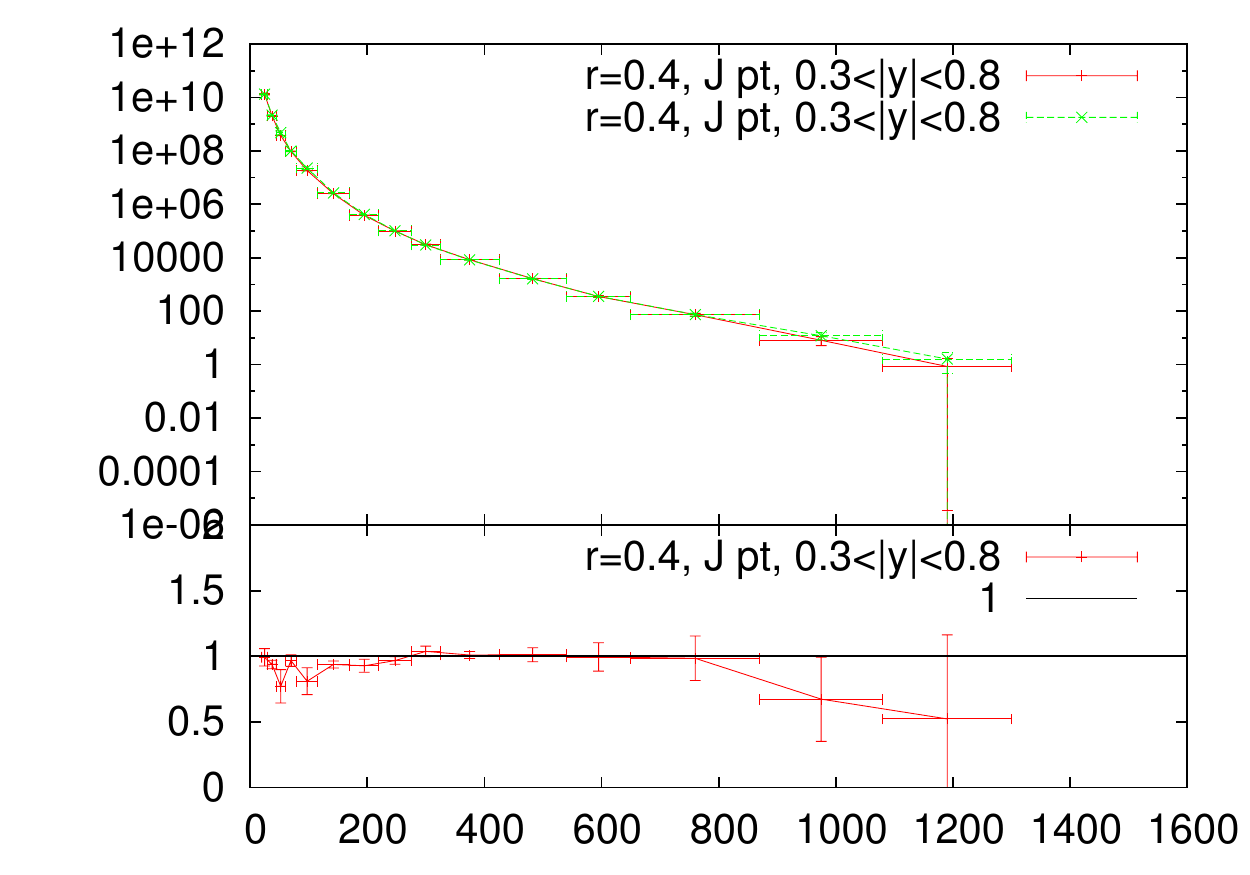,width=0.48\textwidth}
\epsfig{file=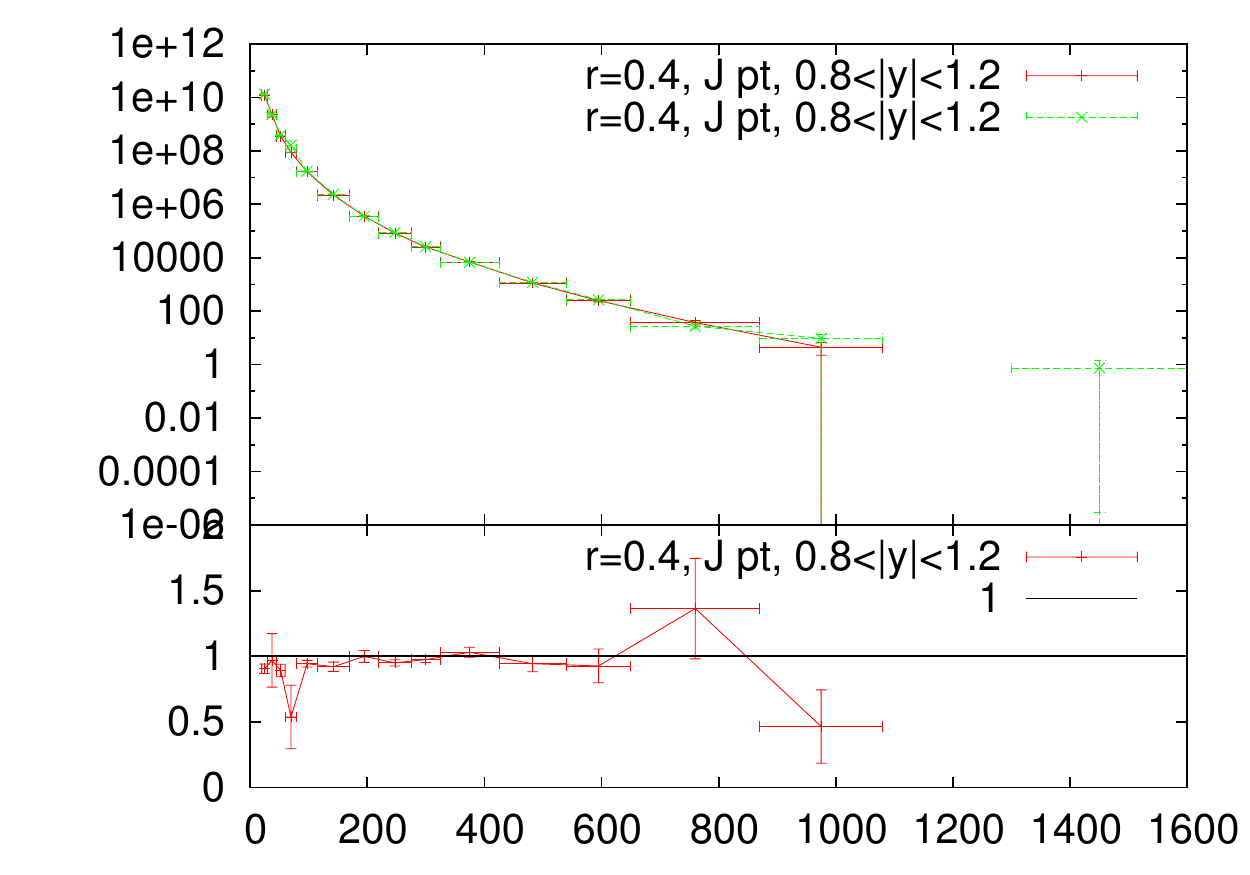,width=0.48\textwidth}
\epsfig{file=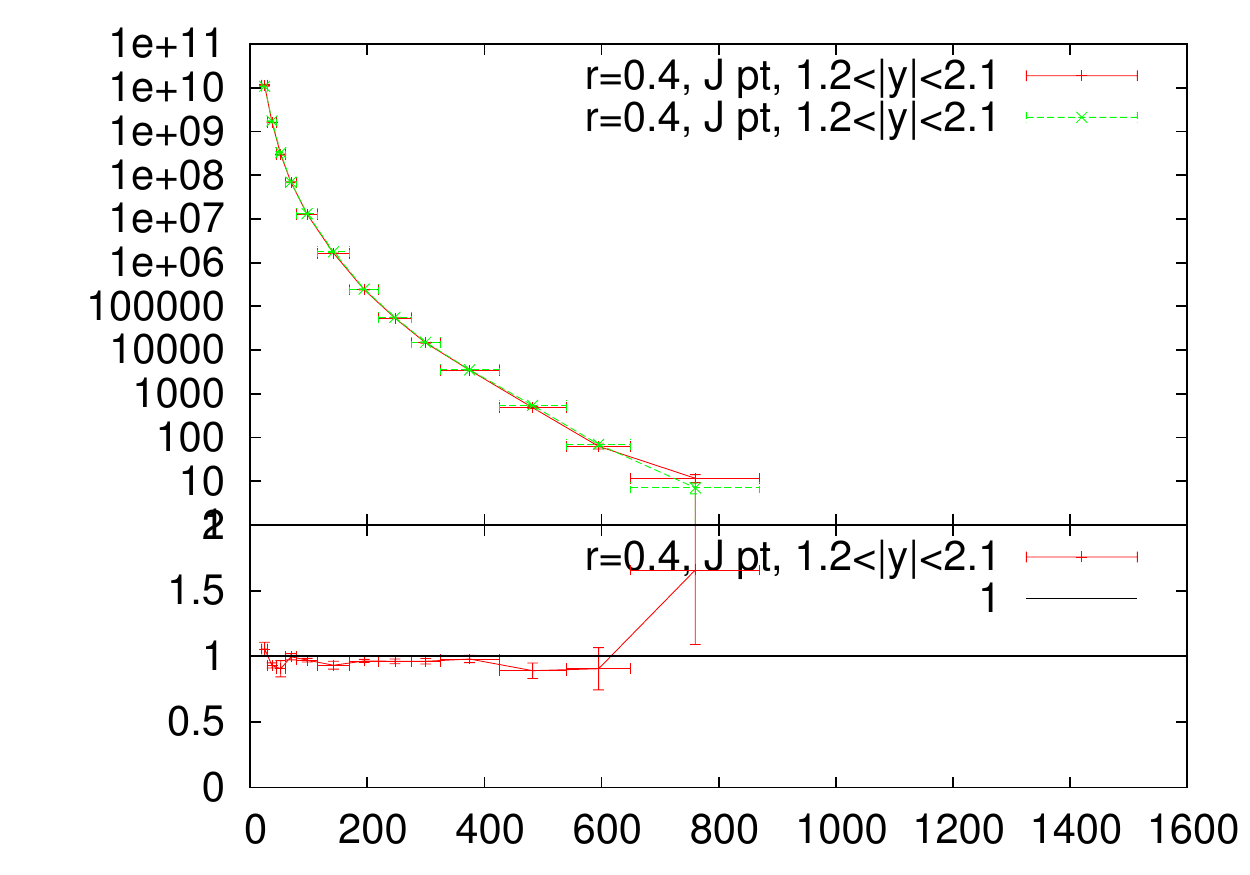,width=0.48\textwidth}
\epsfig{file=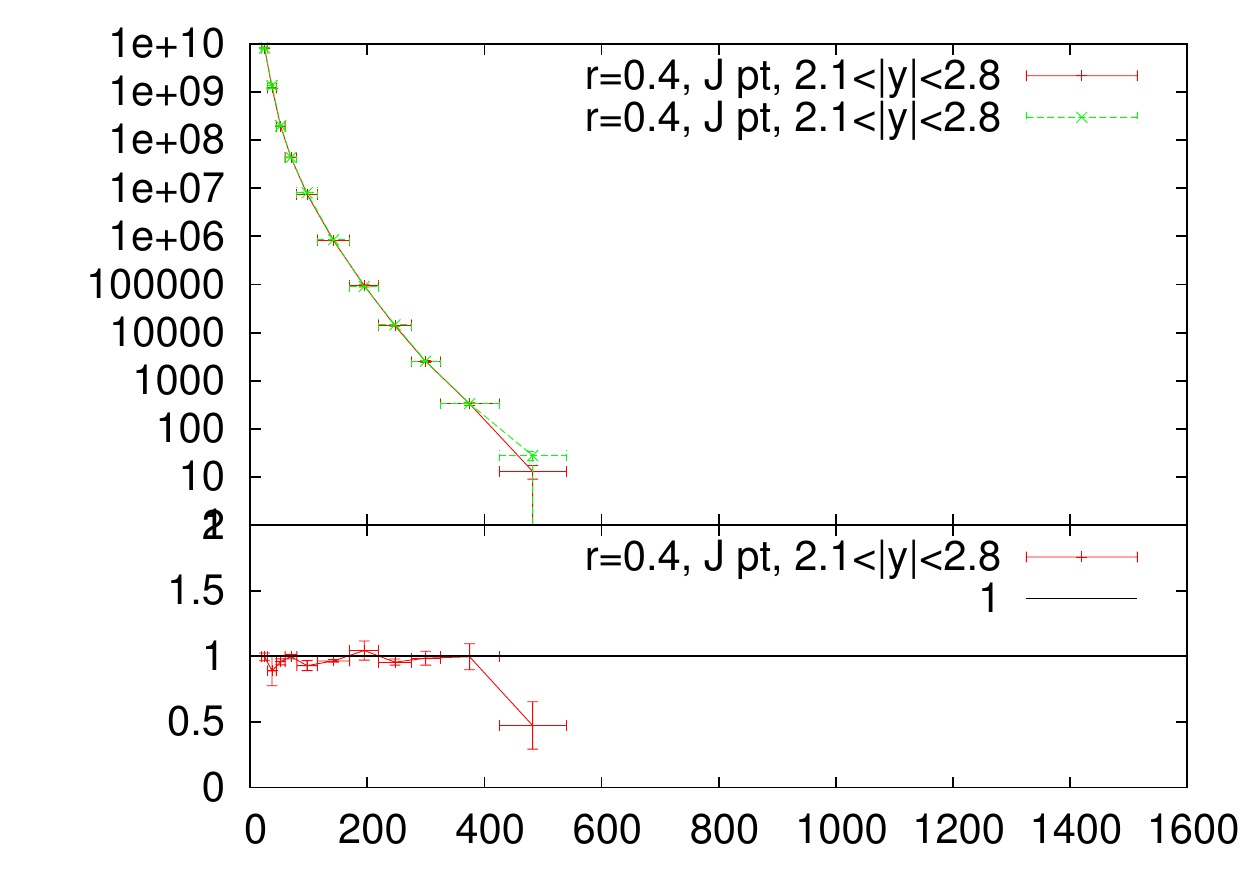,width=0.48\textwidth}
\epsfig{file=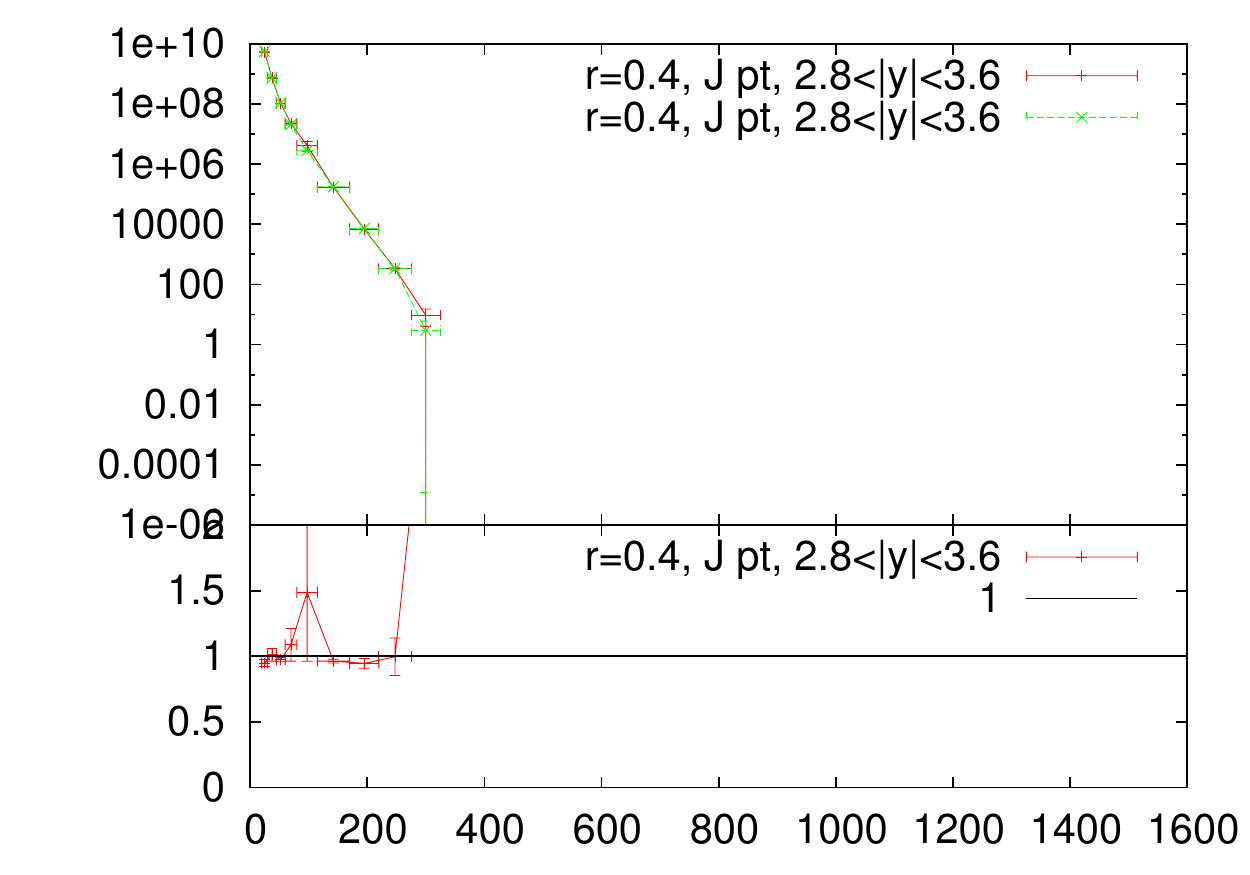,width=0.48\textwidth}
\end{center}
\caption{Comparison of the inclusive jet $\pt$ distribution after shower,
  between the sample obtained with the {\tt doublefsr 1} option and with the
  new separation of regions of ref.~\protect\cite{NasonZanderighi}, and sample
  D1.}
\label{fig:pt8}
\end{figure}

As a final note, we also consider the effect of a further improvement
in the separation of regions in the \POWHEGBOX{}, that has been recently
found to be useful in processes of vector boson production in association
with jets~\cite{NasonZanderighi}. We have generted a sample using
this method, in association with the {\tt doublefsr 1} setting.
The comparison of the result at the Les Houches level versus the showered
one is shown in fig.~\ref{fig:pt7}.
We see a slight improvement, with even less spikes, when the new option is
applied.  On the other hand, no sensible differences are seen between this
new sample and sample D1, as shown in fig.~\ref{fig:pt8}.

\section{Conclusions}
In this note we have introduced a refinement of the way in which the
\POWHEGBOX{} separates the singular regions. This refinement constitutes
without doubt and improvement, although this improvement is only relative to
a small region of phase space.

We have observed that, from a practical point of view, the introduction of
the new feature helps in reducing the appearance of spikes in the computed
distributions. We thus recommend that this feature should be used always in
dijet production.

We have also studied the sensitivity of the results due to a modification of
the way the \POWHEG{} events are passed to the shower Monte Carlo program.
We see no reason why the modified scheme (referred to as RS in the text)
should not be considered as valid as the original \POWHEG{} one.  We thus
conclude that the variation that we have found should be considered an
intrinsic theoretical uncertainty related to NLO+PS matching.  We do not
have, at the moment, a systematic way to assign a theoretical error to this
uncertainty.


\begin{thebibliography}{10}
\bibitem{Alioli:2010xa}
  S.~Alioli, K.~Hamilton, P.~Nason, C.~Oleari and E.~Re,
  JHEP {\bf 1104} (2011) 081
  [arXiv:1012.3380 [hep-ph]].
\bibitem{Sjostrand:2006za}
  T.~Sjostrand, S.~Mrenna and P.~Z.~Skands,
  JHEP {\bf 0605} (2006) 026
  [hep-ph/0603175].
\bibitem{Pumplin:2002vw} 
  J.~Pumplin, D.~R.~Stump, J.~Huston, H.~L.~Lai, P.~M.~Nadolsky and W.~K.~Tung,
  JHEP {\bf 0207}, 012 (2002)
  [hep-ph/0201195].

\bibitem{Cacciari:2008gp}
  M.~Cacciari, G.~P.~Salam and G.~Soyez,
  JHEP {\bf 0804} (2008) 063
  [arXiv:0802.1189 [hep-ph]].

\bibitem{NasonZanderighi}
  J.~Campbell, R.K.~Ellis, P.~Nason and G.~Zanderighi,
  ``$W$ and $Z$ bosons in association with two jets using the POWHEG method'',
  in preparation.

\end{thebibliography}
\end{document}